\documentclass{article}
\usepackage{arxiv}

\usepackage[utf8]{inputenc} 
\usepackage[T1]{fontenc}    
\usepackage{hyperref}       
\usepackage{url}            
\usepackage{booktabs}       
\usepackage{amsfonts}       
\usepackage{nicefrac}       
\usepackage{microtype}      
\usepackage{lipsum}

\usepackage{times}
\usepackage{epsfig}
\usepackage{graphicx}
\usepackage{amsmath}
\usepackage{amssymb}
\usepackage{subfigure}
\usepackage{enumitem}
\usepackage{indentfirst}
\usepackage{multirow}
\usepackage{color}
\usepackage{makecell}

\title{Spatiotemporal Entropy Model is All You Need for Learned Video Compression}

\author{Zhenhong Sun, Zhiyu Tan, Xiuyu Sun\thanks{Corresponding author.}, Fangyi Zhang, Dongyang Li, Yichen Qian, Hao Li\\
	\\Alibaba Group, China\\\\
	\texttt{\{zhenhong.szh, zhiyu.tzy, xiuyu.sxy, zhiyuan.zfy,}\\
	\texttt{yingtian.ldy, yichen.qyc, lihao.lh\}@alibaba-inc.com}
}

\begin{document}
\maketitle

\begin{abstract}
The framework of dominant learned video compression methods is usually composed of motion prediction modules as well as motion vector and residual image compression modules, suffering from its complex structure and error propagation problem. Approaches have been proposed to reduce the complexity by replacing motion prediction modules with implicit flow networks. Error propagation aware training strategy is also proposed to alleviate incremental reconstruction errors from previously decoded frames. Although these methods have brought some improvement, little attention has been paid to the framework itself. Inspired by the success of learned image compression through simplifying the framework with a single deep neural network, it is natural to expect a better performance in video compression via a simple yet appropriate framework. Therefore, we propose a framework to directly compress raw-pixel frames (rather than residual images), where no extra motion prediction module is required. Instead, an entropy model is used to estimate the spatiotemporal redundancy in a latent space rather than pixel level, which significantly reduces the complexity of the framework. Specifically, the whole framework is a compression module, consisting of a unified auto-encoder which produces identically distributed latents for all frames, and a spatiotemporal entropy estimation model to minimize the entropy of these latents. Experiments showed that the proposed method outperforms state-of-the-art (SOTA) performance under the metric of multiscale structural similarity (MS-SSIM) and achieves competitive results under the metric of PSNR.
\end{abstract}


\section{Introduction}

\begin{figure}[t]
	\centering
	\subfigure[Two adjacent frames]{\includegraphics[scale=0.5]{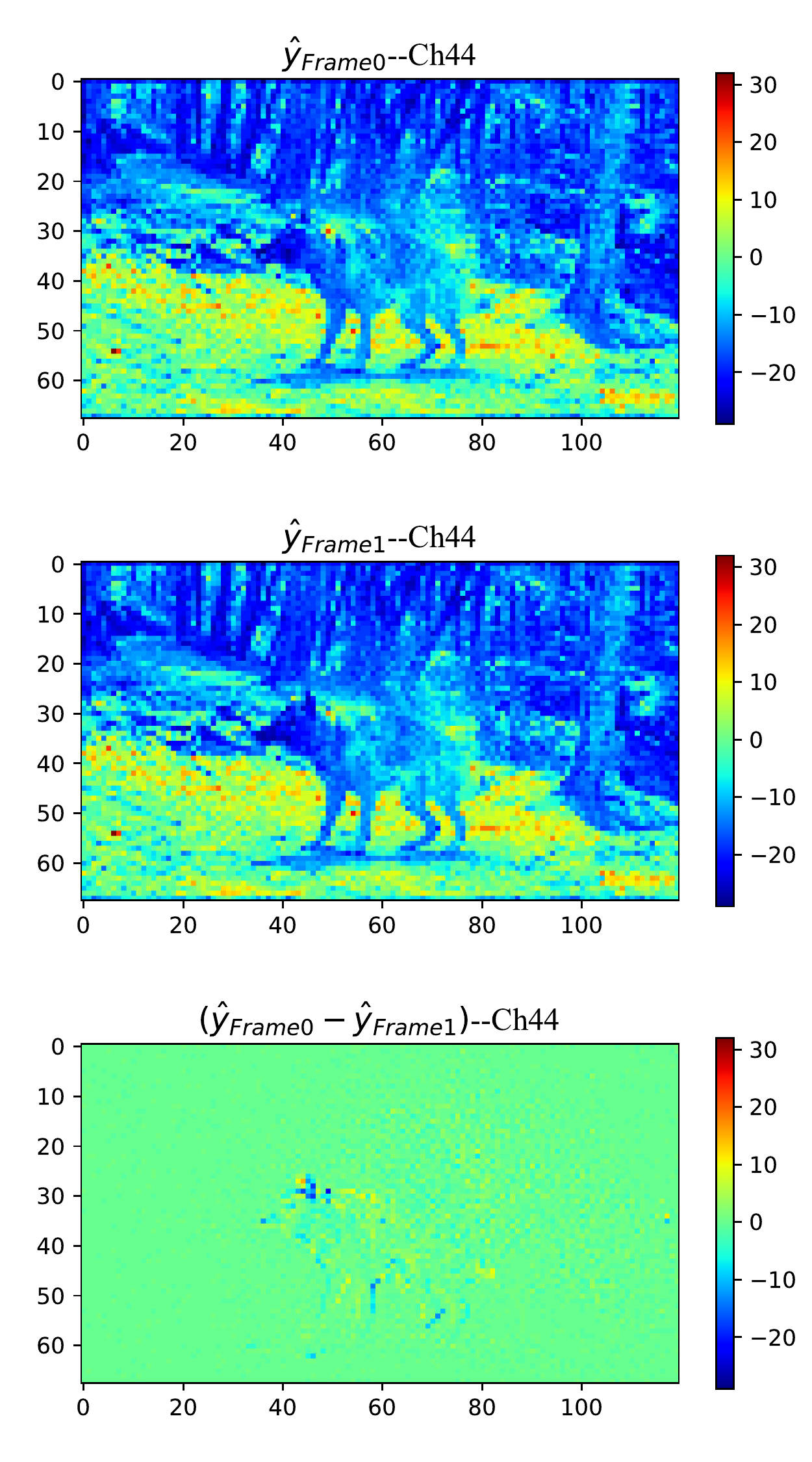}\label{fig:v1}}
	\subfigure[Two separated frames]{\includegraphics[scale=0.5]{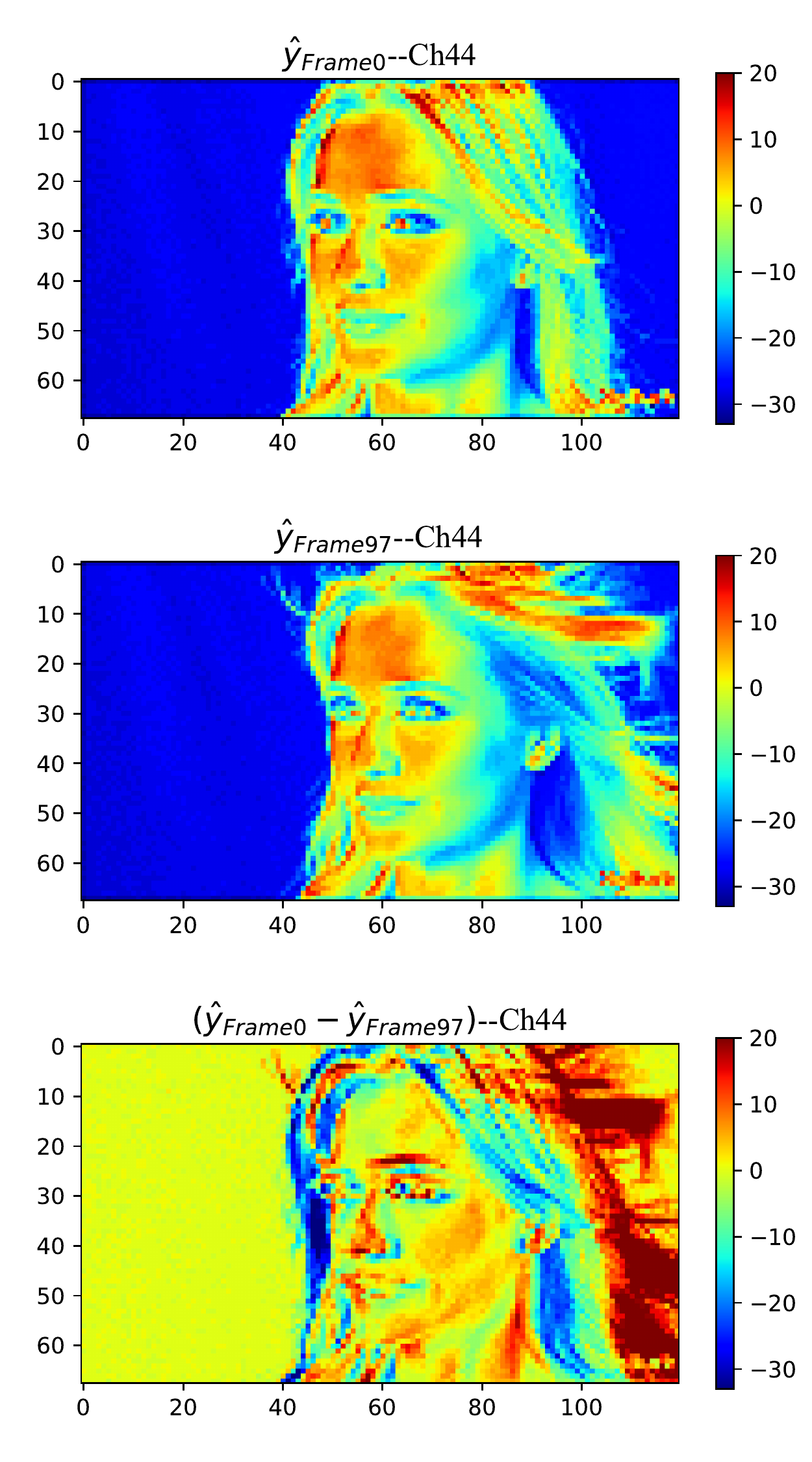}\label{fig:v2}}
	\caption{Correlation between latent representations. The first two rows show the heatmaps of the key channel (channel 44) in latent representations of different frames. The last row shows the differential results of the front two rows. Two adjacent frames in (a) are from the \emph{ShakeNDry} in UVG Dataset, showing high correlations. Two separated frames in (b) are from the \emph{Beauty} in UVG Dataset, showing fewer correlations.}
	\label{fig:visual}
	\vspace{-0.4cm}
\end{figure}

In the field of video compression, eliminating temporal redundancy between adjacent frames is a key challenge to improve compression performance~\cite{avc,hevc,vvc}. Traditional video standards such as H.264~\cite{avc}, HEVC~\cite{hevc} and VVC~\cite{vvc} use well hand-designed modules (\emph{e.g.}, block motion estimation, motion vector and residual image transform) to reduce the redundancy in video sequences.

Under the traditional video coding schemes, some successful attempts~\cite{dvc,dvc2,djelouah2019interp,hlvc,mlvc,liuhaojie2019,google2020} for learned video compression replace each module with a learning-based one. 
Specially, the recursive block-based motion prediction is replaced by optical flow with warping operation to generate predictive frames, and the modules to compress residual errors and motion vector are replaced by two auto-encoders styled compression networks.
To further exploit the temporal redundancy, multi-frame prediction~\cite{djelouah2019interp,hlvc,mlvc} and implicit optical flow methods~\cite{liuhaojie2019,google2020} are proposed to improve the performance of motion prediction.

The aforementioned learned video compression methods have shown competitive performance to traditional codecs, but they still suffer from the following three inherent problems caused by their framework:
\begin{itemize}[leftmargin=15pt,itemsep=2pt,topsep=2pt]
\item[$\bullet$]  Due to the mechanism of using a previous reconstructed frame as reference, error propagation problem is common in the methods based on the traditional framework, for no matter codecs or learned methods~\cite{dvc2}; 
\item[$\bullet$]  The accuracy of a predicted frame may decrease due to errors from motion prediction or propagated from earlier frames, resulting in residual errors which will then increase the difficulty of residual compression;
\item[$\bullet$]  Training and inference with the current framework are computationally expensive due to its complexity. 
\end{itemize}

Therefore, it comes to an open question: \emph{Does learned video compression method really need such a complex framework?} At first glance, the answer seems to be of course \emph{YES}. Obviously, the advance of this framework has been proved over the last decades. However, we have a different view in this paper. Our inspiration comes from the development of learned image compression: some learning-based methods achieved more remarkable performance with only a simple yet powerful entropy model~\cite{balle2017variational,theis2017lossy,balle2018variational,minnen2019joint} without following the advanced image codec schemes. 
Following this phenomenon, it is natural to expect that a similar simple framework (\emph{e.g.}, an appropriate spatiotemporal entropy model) may be also good enough for learned video compression. 

To verify this intuition, latent representations from a unified image compression model are visualized for adjacent frames, as shown in Fig.~\ref{fig:visual}, from which we can observe obvious differences highly correlated to temporal redundancy. Motivated by this, we propose a motion-free video compression (MFVC) framework. 
To be specific, we view frames in a video sequence as independent pictures in the encoding process and use an encoder network to project each frame into quantized latent representations respectively, similar to that in learned image compression methods. Based on these latent representations, a spatial prior module (SPM) is then used inside frames, and a temporal prior module (TPM) is used over adjacent frames, jointly providing accurate entropy for a common arithmetic encoding. In the decoding process, after arithmetic decoding, the lossless quantized latents are projected back to image pixels. 
There exists no lossy reference-frame dependence, no motion prediction, and no motion and residual compression in the whole scheme.

In particular, this paper has three major contributions:
\begin{itemize}[leftmargin=15pt,itemsep=2pt,topsep=2pt]
	\item[$\bullet$] To the best of our knowledge, the motion-free video compression framework is the first learned video compression framework without motion prediction modules while achieving SOTA performance.
	\item[$\bullet$] We propose the spatiotemporal entropy model for video frame latent representation compression, which consists of joint hyper-prior encoder-decoder (P-HE/P-HD), spatial prior module, and temporal prior module. 
	\item[$\bullet$] Experimental results reveal that the motion-free video compression framework achieves SOTA performance under the MS-SSIM and firstly realizes the variable-rate control in a single video compression model.
\end{itemize}


\section{Related Work}
\subsection{Learned Image Compression}
Various handcrafted-based image compression standards have been proposed over the past decades, such as ~\cite{jpeg,jpeg2000,bpg}. 
The newest hand-designed image compression methods are developed from video standards. With block partitioning, intra prediction, residual compression, and other modules, they require about half the storage space as the equivalent quality JPEG (including fixed block partition, DCT/IDCT transform, and Huffman Coding). 

Recently, some learning-based lossy image compression approaches have been proposed to achieve competitive performance~\cite{toderici2015variable,toderici2017full,theis2017lossy,balle2017variational,balle2018variational,johnston2018improved,minnen2019joint,lee2019,cheng2020learned,lee2019hybrid}. They all utilize auto-encoder style networks to transform pixels to quantized latent representations and then project back to the pixel space. Indeed, a jointly optimized entropy model is verified to make the latent presentations more suitable to be losslessly compressed to a bitstream based on entropy coding. Some works~\cite{balle2017variational,toderici2017full,lee2019hybrid} try to improve the representation ability of the auto-encoder architecture. Ball{\'e} \emph{et al.}~\cite{balle2017variational} bring in the generalized divisive normalization (GDN) transform with optimized parameters to efficiently Gaussianize the local joint statistics of natural images. Toderci \emph{et al.}~\cite{toderici2017full} introduce a new gated recurrent unit (GRU) inspired by the residual network. Cheng \emph{et al.}~\cite{cheng2020learned} introduce deep residual attention modules to improve the performance. 

More works~\cite{balle2017variational,balle2018variational,toderici2017full,minnen2019joint,lee2019,qian2020learning} pay attention to the entropy network and obtain obvious improvements. Ball{\'e} \emph{et al.}~\cite{balle2017variational} use a fully factorized prior to minimize the entropy of the elements of the whole latent representation. An improved version~\cite{balle2018variational} is proposed to use a hierarchical learned prior network to occupy the fact that spatially neighboring elements of the latent representation tend to vary together in their scales. Furthermore, context-adaptive based models~\cite{toderici2017full,minnen2019joint,lee2019,lee2019hybrid} are introduced to utilize a neural network like PixelCNN~\cite{pixelcnn} to incorporate predictions from neighboring symbols, avoiding storing additional bits. This kind of method outperforms the top standard image codec (BPG) on both the PSNR and MS-SSIM distortion metrics. 
After that, Qian~\cite{qian2020learning} builds up a global relevance throughout latent features to further explore the relationship in the whole picture. All these works show that designing an accurate entropy model is crucial or even the only thing for learned image compression. 

Last but not the least, rate control in learning-based methods is different from the codec schemes too. Early learning-based methods always need to train different models for different reconstruction quality. Toderci \emph{et al.}~\cite{toderici2017full} use recurrent neural networks to recursively compress residual information similar to the recursive search block in video codecs, encoding the latents of each iteration by a binary representation. Choi \emph{et al.}~\cite{choi2019variable} propose a conditional convolution to make rate control easier in auto-encoder style networks without incremental inferences.

\subsection{Learned Video Compression}
Thanks to the success of learned image compression, more attempts are verified to see where learning can obtain gain in video compression~\cite{wu2018video,3D2019,3D2020,dvc,dvc2,djelouah2019interp,liuhaojie2019,hlvc,mlvc,google2020}. DVC~\cite{dvc} is proposed by replacing each module for the traditional codec as a CNN network. All of these networks are trained in an end-to-end manner and optimized by PSNR or MS-SSIM loss functions. It can be seen as a "deep" version of traditional video coding scheme: an optical flow network (pre-trained spynet~\cite{spynet}) is utilized to predict motion between a frame and its previous compressed frame; a motion compensation network is applied over warped frames as a learning-based loop filter; finally, motion and residual information are compressed by two auto-encoder style compression networks. 
Bidirectional motion prediction method~\cite{djelouah2019interp} and multi-frames based unidirectional one~\cite{mlvc} obtain advantages over DVC in general. Hierarchical learned video compression (HLVC) ~\cite{hlvc} extends these methods by combining unidirectional and bi-directional compression with a weighted recurrent quality enhancement network. These methods are all based on the traditional video coding scheme which has some inherent problems: error propagation owing to the use of lossy frames (from lossy reconstruction) as references, inefficient residual compression caused by inaccurate motion prediction, too complicated networks to convergence due to the extra calculations and complicated modules in pre-trained optical flow models. 

To tackle these problems, some incremental solutions are introduced. Lu \emph{et al.}~\cite{dvc2} firstly point out the error propagation problem in learned video compression. They design a joint training strategy to train the video codec by using the information from different time steps in one video clip and combines all the information to optimize the learned codec for better video compression performance. This method can alleviate but not avoid the error propagation problem. In order to reduce the failure cases (\emph{e.g.}, the disocclusion and fast motion cases) derived from the optical flow and bilinear warping operation, a scale-space flow with scale-space warping operation is introduced in~\cite{google2020}, which achieves the SOTA performance among other learned methods. Implicit optical flow based methods~\cite{liuhaojie2019,google2020} use an encoder network to directly aggregate motion information instead of a pre-trained flow network, reducing the cost of model storage, computation, and training. 

Some researchers try to not use the "traditional" learned video compression framework. Wu \emph{et al.}~\cite{wu2018video} view video compression as an image interpolation problem and propose to interpolate target frame from references in a hierarchical manner, where the residual of the interpolation is then compressed via a RNN-based image compression network. Newest works~\cite{3D2019,3D2020} try to directly capture spatiotemporal relationships using a 3D auto-encoder and autoregressive prior model. These methods outperform H264, but worse than HEVC and other learned methods.

\begin{figure}[t]
	\centering
	\includegraphics[scale=0.7]{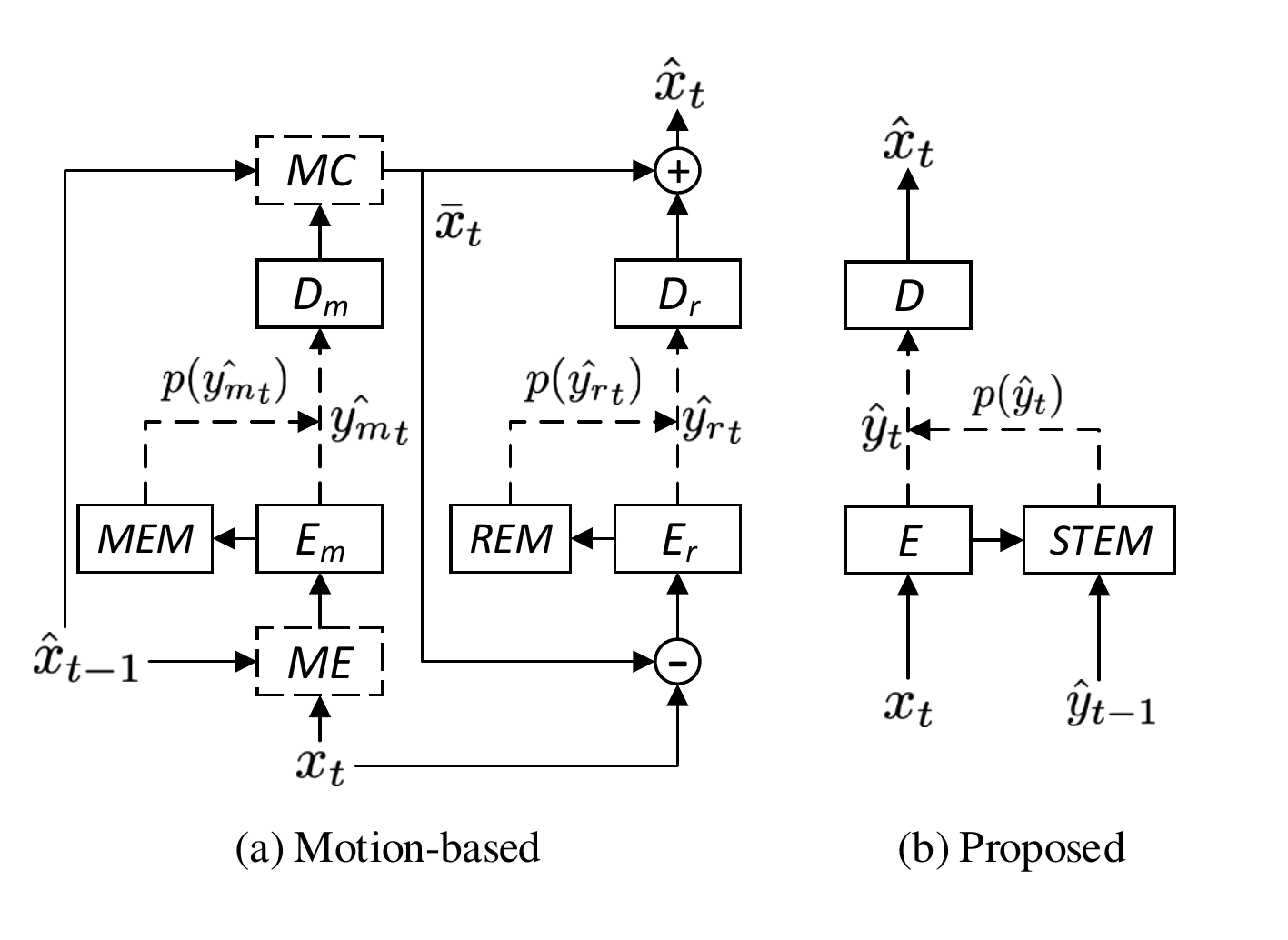}
	\caption{Operational diagrams of motion-based video compression framework (a) and the proposed framework (b). MEM: Motion Entropy Model, ME: Motion Estimation (\emph{e.g.}, optical flow), $E_m$: Motion Encoder, $D_m$: Motion Decoder, MC: Motion Compensation (\emph{e.g.}, wraping operation and post-processing network), REM: Residual Entropy Model, $E_r$: Residual Encoder, $D_r$: Residual Decoder, $E$: Encoder, $D$: Decoder, STEM: Spatiotemporal Entropy Model. The motion prediction consists of ME and MC. $x_t$, $\bar{x}_t$, $\hat{x}_{t-1}$, and $\hat{x}_{t}$ represent a current frame, predictive frame, previous reconstructive frame, current reconstructive frame, respectively. $\hat{y}_{..}$ and $p(\hat{y}_{..})$ represent the quantitative latent and the probability distribution, respectively.}
	\label{fig:IPC1}
	\vspace{-0.4cm}
\end{figure}

The 3D convolution idea provides new thinking on video compression, \emph{i.e.}, a simple framework like that for learned image compression might be good enough for learned video compression, but building a unified spatiotemporal transform is much more complex and challenging than the spatial one, particularly with the 3D convolution which requires multiple frames as inputs. In addition, too computational expensive models like the 3D auto-encoder are also not suitable for low-delay scenarios. 
To this end, in this paper, we propose a motion-free video compression with a spatial transform and a spatiotemporal entropy model. The spatial transform keeps the advantages coming from learned image compression, and the spatiotemporal entropy model eliminates temporal redundancy in a lossless manner between frames without breaking the transformed representation. 

\section{Proposed Method}
\subsection{Motion-based Video Compression Framework}
\subsubsection{Complexity}\label{sec:311}
The mainstream framework of learned video compression consists of many networks: motion prediction network, motion vector compression network, and residual image compression network, as shown in Fig.~\ref{fig:IPC1}.
Through the motion estimation network (generally, optical flow network), motion vector $mv$ is predicted between previous reconstructed frame $\hat{x}_{t-1}$ and current frame $\hat{x}_t$, and then it will be compressed with motion vector compression network.
After that, the predictive frame $\bar{x}_{t}$ is generated based on the reconstructed motion vector $\hat{mv}$ with warping operation from previous reconstructed frame $\hat{x}_{t-1}$. 
Finally, residual error $x_r$ between $\bar{x}_{t}$ and $x_t$ is compressed with residual image compression network. 
Additionally, each compression network is along with the entropy model networks to efficiently model the latent representations $\hat{y}$ for more accurate entropy minimization. 

In the encoding phase, motion decoding, motion compensating, and residual decoding are required to compute the previous reconstructed frame $\hat{x}_{t-1}$, which results in a very high computational cost. 
Although the explicit optical flow network is omitted in~\cite{liuhaojie2019, google2020} to reduce the model storage and computation, the framework with two compression networks and entropy models networks is still complicated for practical applications, considering the model storage, computation, and variable-rate control.

\vspace{-0.3cm}
\subsubsection{Error Propagation problem}
Error propagation between the reconstructed frames is a common problem in both traditional and learned video compression methods~\cite{dvc2}. 
To investigate details of the error propagation, multiple frames are compressed by two methods (\emph{i.e.}, H.264 and flow-based learned video compression).
The PSNR values of the reconstructed frames are demonstrated in Fig.~\ref{fig:error}, and the values of these methods are high at the beginning of the group of piture~\cite{gop} (GOP). 
While, as the GOP size increases, the PSNR values begin to drop from 35.24 dB to 31.05 dB for H.264 and from 34.41 dB to 31.324 dB for the flow-based method, which indicates the error propagation problem causes frames quality to degrade.

\begin{figure}[t]
	\centering
	\includegraphics[scale=0.5]{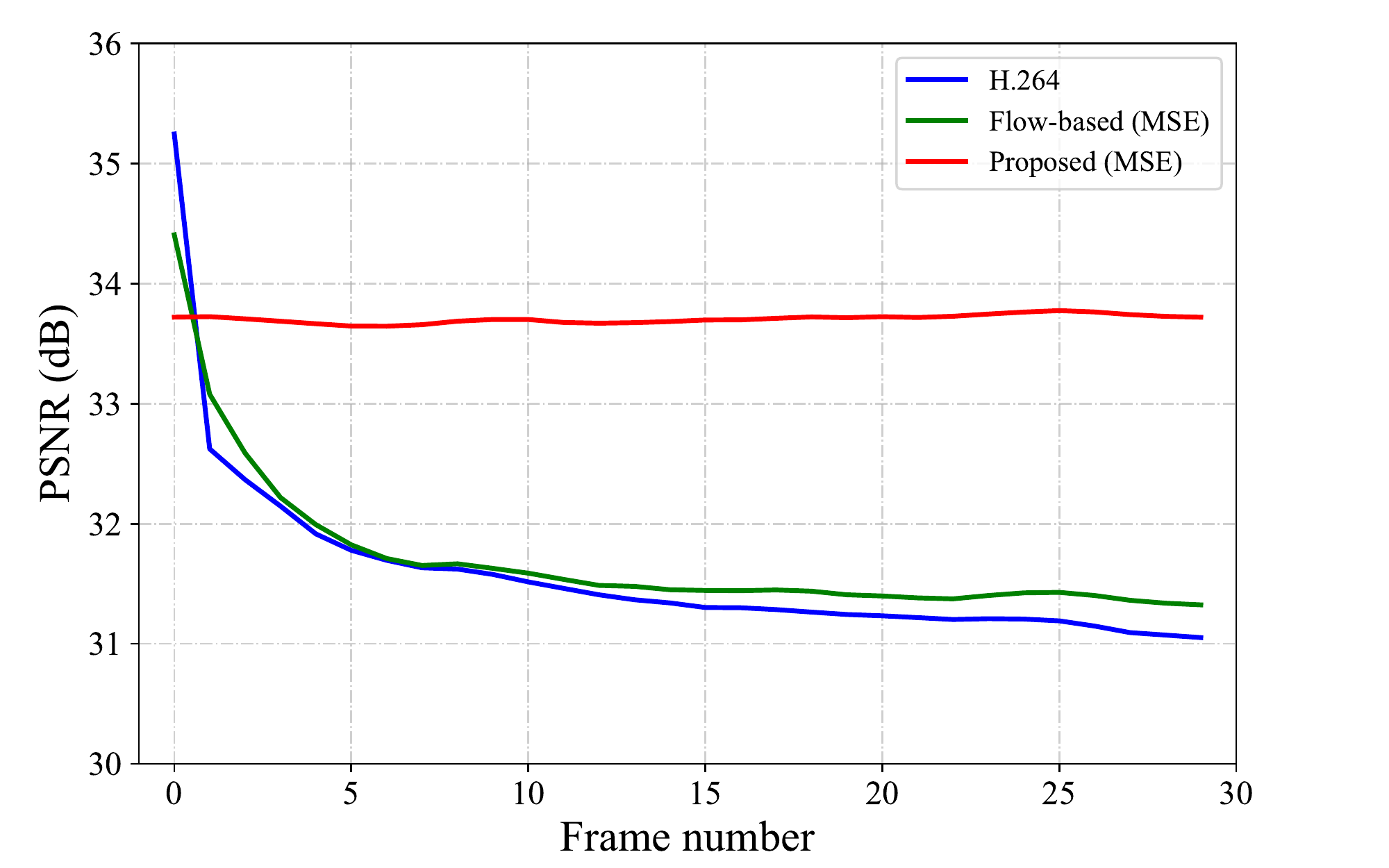}
	\caption{PSNR values for multi-frames compression on HEVC Class B dataset. H. 264 is conducted at a fixed compression rate (\emph{i.e.}, quantization parameter of 27). Motion-based video compression framework is a typical learned video compression with an optical flow and a wraping operation in a single rate model.}
	\label{fig:error}
	\vspace{-0.4cm}
\end{figure}

\begin{figure*}[t]
	\centering
	\includegraphics[scale=0.45]{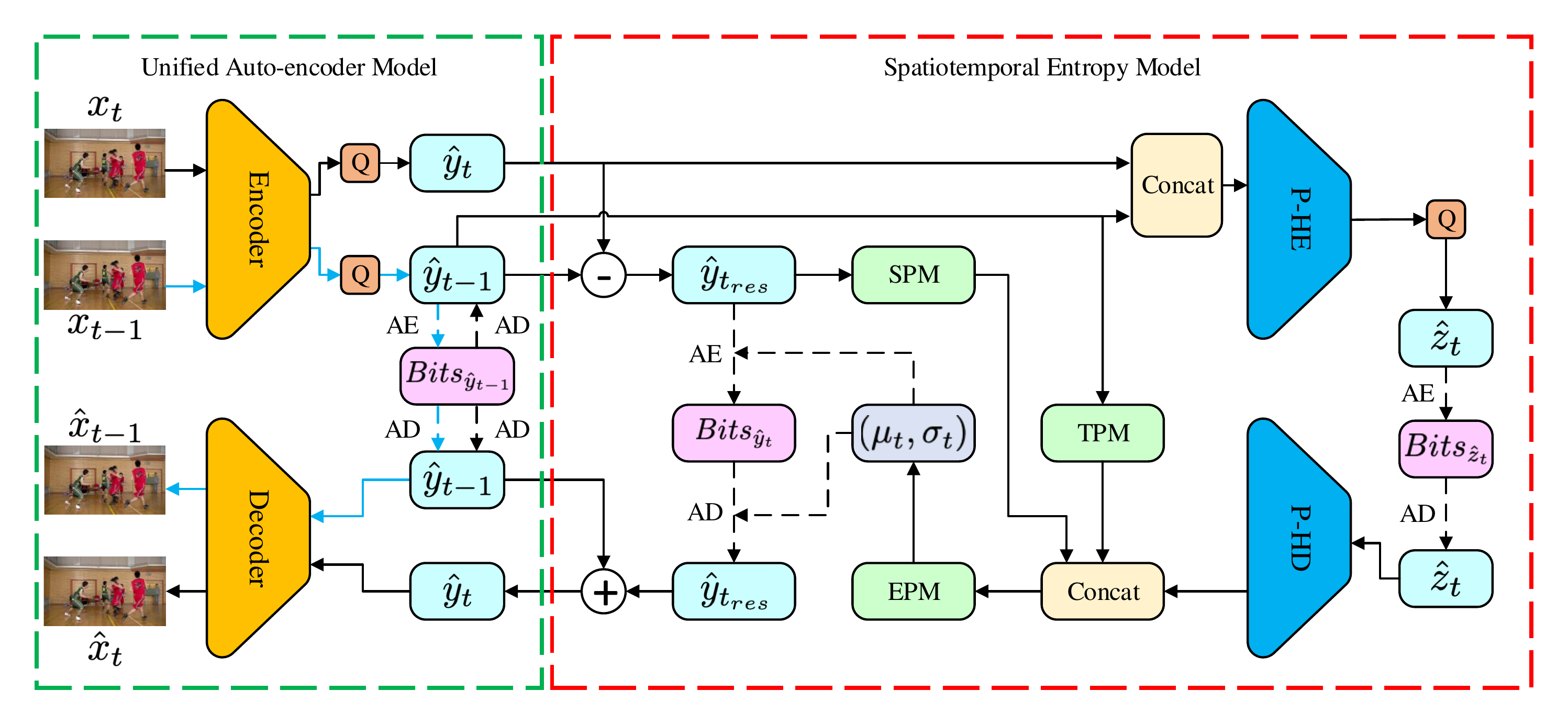}
	\caption{MFVC Framework for low-delay scenarios. Q: Quantization, SPM: Spatial Prior Module, TPM: Temporal Prior Module, EPM: Entropy Parameter Module, P-HE: P-frame Hyper-prior Encoder, P-HD: P-frame Hyper-prior Decoder, AE/AD: Arithmetic Encoding/Decoding. The dashed line indicates the AE/AD. The blue line indicates the previous frame ($x_{t-1}$) compression without the entropy network (I-frame or P-frame). The black line indicates the current P-frame ($x_t$) compression. In the encoding of $x_t$, $\hat{y}_{t-1}$ is obtained from the previous frame ($x_{t-1}$) compression, which puts the $\hat{y}_{t-1}$ in the buffer. In the decoding of $\hat{x}_t$, $\hat{y}_{t-1}$ is decoded from $Bits_{\hat{y}_{t-1}}$ by AD.}
	\label{fig:IPC2}
	\vspace{-0.4cm}
\end{figure*}

Due to the distortion of the predictive frame, which is caused by warped from the previous reconstructed frame, it will generate high-frequency residual regions. 
Thus, the compression of high-frequency regions is difficult for a residual compression network. 
Therefore, if the codec does not increase the bitstream for the residual, the error will propagate to the subsequent frames, and PSNR values will drop continuously with the increasing time steps. 
If increasing the bitstream continuously, the RD performance will degrade and the rate control will lose efficacy.
Although inserting high-quality frames in traditional video compression and utilizing the error propagation aware strategy in learned video compression~\cite{dvc2} help to alleviate the distortion, the error propagation cannot be completely evitable for the motion-based video compression framework. 
Instead, the proposed motion-free video compression framework maintains the same quality level along with the frames (\emph{i.e.}, \emph{Proposed} in Fig.~\ref{fig:error}), and the new framework will be explained in the next subsection.

\subsection{Motion-free Video Compression Framework}
As shown in Fig.~\ref{fig:visual}, the \emph{ShakeNDry} sequences show that a dog throwing water on its body in Fig.~\ref{fig:v1}, showing high correlations between two adjacent frames, and the \emph{Beauty} sequences show that a lady slowly turns her head and wriggles her hair in Fig.~\ref{fig:v2}, showing fewer correlations between two separated frames. 
Base on the observation, the temporal redundancy between frames can be reduced in latent space by minimizing the entropy. 
To this end, the motion-free video compression framework is proposed.
A unified auto-encoder model is used for mapping intra-frames and inter-frames into latent representations (which are all under the same distributions), and then a spatiotemporal entropy model is proposed to estimate the entropy of current-frame latent with the reference-frame latent in the inter-frames compression.

A practical diagram of the MFVC framework is displayed in Fig.~\ref{fig:IPC2}, and this structure is mainly designed for low-delay scenarios (P-frame compression) without considering the bidirectional scenarios (B-frame). 
The basic compression network is a unified auto-encoder model (similar to~\cite{minnen2019joint}) for transforming all input frames $x$ into latents ${\hat{y}}$ and restoring the reconstructed frames ${\hat{x}}$ from the latents. 
Latents can be losslessly compressed by arithmetic encoding (AE)~\cite{arithmatic} and transmitted into a string of bits, using a probability distribution $p_{\hat y}(\hat{y})$. 
When processing the first frame $x_0$ of each GOP, the probability distribution $p_{\hat{y}_0}(\hat{y}_0)$ is generated by a spatial entropy model same as that in image compression methods.
The probability distributions of other frames are all estimated by the spatiotemporal entropy model. In this work, the spatiotemporal entropy model and unified auto-encoder model are trained separately. 

For the sake of simplicity, we take the $t$-th frame compression as an example to explain the P-frame compression process without considering I-frame compression. 
$\hat{y}_t$ and $\hat{y}_{t-1}$ are generated from input frames ($x_{t}$ and $x_{t-1}$) by the fixed Encoder network, downsampling with 16x size scaling. 
To further reduce the information entropy, we use the differential $\hat{y}_{t_{res}}$ between $\hat{y}_t$ and $\hat{y}_{t-1}$ as the input of AE. 
To enable more accurate probability distribution, the temporal compaction is conducted using three prior modules (\emph{e.g.}, hyper-prior Encoder-Decoder, spatial prior module, and temporal prior module) with $\hat{y}_t$ and $\hat{y}_{t-1}$ as inputs. 
The P-HE network takes $\hat{y}_t$ and $\hat{y}_{t-1}$ as inputs to generate the hyper latent $\hat{z}_{t}$, the SPM network uses PixelCNN structure~\cite{pixelcnn} to capture the spatial priors between the neighbors in $\hat{y}_{t_{res}}$, and the TPM network uses three convolutions with Leaky ReLU to extract temporal priors from the previous latent $\hat{y}_{t-1}$. 
To generate the probability distribution parameters $({\mu_t}, {\sigma_t})$, the EPM module is introduced to fuse the outputs of P-HD, SPM and TPM. 
Similar to~\cite{zhoulei}, the probability $p_{\hat y}(\hat{y})$ of quantized latent ${\hat y}$ is modeled as Laplacian distribution: 
\begin{equation}
	\setlength{\abovedisplayskip}{3pt}
	\setlength{\belowdisplayskip}{3pt}
	p_{\hat{y}_{t_{res}}}(\hat{y}_{t_{res}}|{\hat{y}_{t-1}}, {\hat{z}_{t}})=\prod_{i=1}\Big(\int_{{\hat{y}_i}-\frac{1}{2}}^{{\hat{y}_i}+\frac{1}{2}} Lap({\hat{y}_{t_{res}}};\mu_t, e^{\sigma_t})\, dy\Big).
	\label{eq:py1}
\end{equation}
For the hyper latent $\hat{z}_{t}$, a set of channel-wise trainable parameters $(\mu_{z_{t}}, \sigma_{z_{t}})$ are defined to represent the Laplacian distribution:
\begin{equation}
	\setlength{\abovedisplayskip}{3pt}
	\setlength{\belowdisplayskip}{3pt}
	p_{\hat{z}_{t}}(\hat{z}_{t})=\prod_{i=1}\Big(\int_{{\hat{z}_i}-\frac{1}{2}}^{{\hat{z}_i}+\frac{1}{2}} Lap(\hat{z}_{t};\mu_{z_{t}}, e^{\sigma_{z_{t}}})\, dz\Big).
	\label{eq:pz1}
\end{equation}

The latent $\hat{y}_{t_{res}}$ and the hyper latent $\hat{z}_{t}$ are both compressed by AE and AD, the cost of transmitting $\hat{z}$ is included into the loss function of the spatiotemporal entropy model: 
\begin{equation}
	\setlength{\abovedisplayskip}{3pt}
	\setlength{\belowdisplayskip}{3pt}
	\begin{split}
	\mathbb{L}_P &= R_{\hat{y}_{t_{res}}}+R_{\hat{z}_{t}}\\
	&=\mathbb{E}_{x\sim p_x}[-\log_2{p_{\hat{y}_{t_{res}}}(\hat{y}_{t_{res}})}]+\mathbb{E}_{x\sim p_x}[-\log_2{p_{\hat{z}_{t}}({\hat{z}_{t}})}].
	\end{split}
	\label{eq:loss3}
\end{equation}
Since the quality of the reconstructed frame is only determined by the unified auto-encoder model, the reconstructed distortion of the P-frame is left out of the loss function $\mathbb{L}_P$, simplifying the training process. 
Based on the MFVC framework, it does not require motion compensating, motion and residual decoding in the encoding process, which reduces the storage and computation of the model. 

\begin{table*}[t]
	\begin{center}
		\renewcommand\arraystretch{1.}
		\setlength{\abovecaptionskip}{0pt}%
		\setlength{\belowcaptionskip}{10pt}%
		\scalebox{0.9}{
			\begin{tabular}{|c|c|c|c|c|}
				\hline
				P-HE & P-HD  & TPM  & SPM  & EPM \\
				\hline
				Conv: $3\times3$ c256 s1 & Deconv: $5\times5$ c256 s2 & Conv: $5\times5$ c426 s1 & Masked: $5\times5$ c640 s1 & Conv: $1\times1$ c1600 s1\\
				Leaky ReLU 						 & Leaky ReLU 							& Leaky ReLU 						 & 											   & Leaky ReLU \\
				Conv: $5\times5$ c256 s2 & Deconv: $5\times5$ c256 s2 & Conv: $5\times5$ c533 s1 &											    & Conv: $1\times1$ c1280 s1\\
				Leaky ReLU 						 & Leaky ReLU 							& Leaky ReLU						 & 											   & Leaky ReLU \\
				Conv: $5\times5$ c256 s2 & Conv: $3\times3$ c640 s1   & Conv: $5\times5$ c640 s1  &												 & Conv: $1\times1$ c640 s1\\
				\hline
		\end{tabular}}
	\end{center}
	\caption{Details of each module in the P-frame compression network.
		The "Conv" prefix corresponds to convolutional layers followed by the kernel size, the number of output channels, and downsampling stride (\emph{e.g.}, the first layer of the P-HE uses $5\times5$ kernels with 256 channels and a stride of one). 
		The "Deconv" prefix corresponds to upsampled convolutions (\emph{i.e.}, in TensorFlow, tf.conv2d\_transpose).
		The "Masked" prefix corresponds to a masked convolution as in~\cite{pixelcnn}.}	
	\label{table:IPC2}
	\vspace{-0.4cm}
\end{table*}

\vspace{-0.2cm}
\subsection{Variable Rate Control}
In learned image compression methods, optimizing the network with different Lagrange multiplier ($\lambda$) can trade off the distortion against the rate~\cite{balle2017variational,theis2017lossy} and the learning goal of these methods can be formulated as:
\begin{equation}
	\setlength{\abovedisplayskip}{3pt}
	\setlength{\belowdisplayskip}{3pt}
	\begin{split}
	\mathbb{L}_I &= R+\lambda D \\
	&\textup{with}\space\lambda\in\{\lambda_1, \lambda_2, ..., \lambda_{n-1}, \lambda_{n}\},
	\end{split}
	\label{eq:loss1}
\end{equation}
where $\lambda$ corresponds to $n$ different compression rates, $R$ is referred to as the expected length of the compressed bitstream, and $D$ is the expected distortion of the reconstructed image concerning the original image measured by either mean squared error (MSE) or MS-SSIM. 
Drawing lessons from~\cite{choi2019variable}, conditional convolutions are added into the autoencoder network~\cite{minnen2019joint} for the variable-rate control in the unified auto-encoder model. 
Apart from the input frame, the unified auto-encoder model also takes $\lambda$ as input to enable the conditional convolutions and produces a compressed image with varying rates.

Benefitting from transferring the temporal compaction from the pixel space to the latent space, the loss function $\mathbb{L}_P$ only concentrates on the compressed bitstream of the latent $\hat{y}_{t_{res}}$  and the hyper latent $\hat{z}_{t}$. When training the spatiotemporal entropy model, the latents $\hat{y}_{t}$ and $\hat{y}_{t-1}$ are produced by the fixed unified auto-encoder model using randomly selected $\lambda$ from the set $\{\lambda_1, \lambda_2, ..., \lambda_{n-1}, \lambda_{n}\}$. Based on this strategy, the P-frame compression can achieve the variable-rate control directly without any other operations.

\section{Experiments}

\begin{figure*}[t]
	\centering
	\subfigure{\includegraphics[scale=0.25]{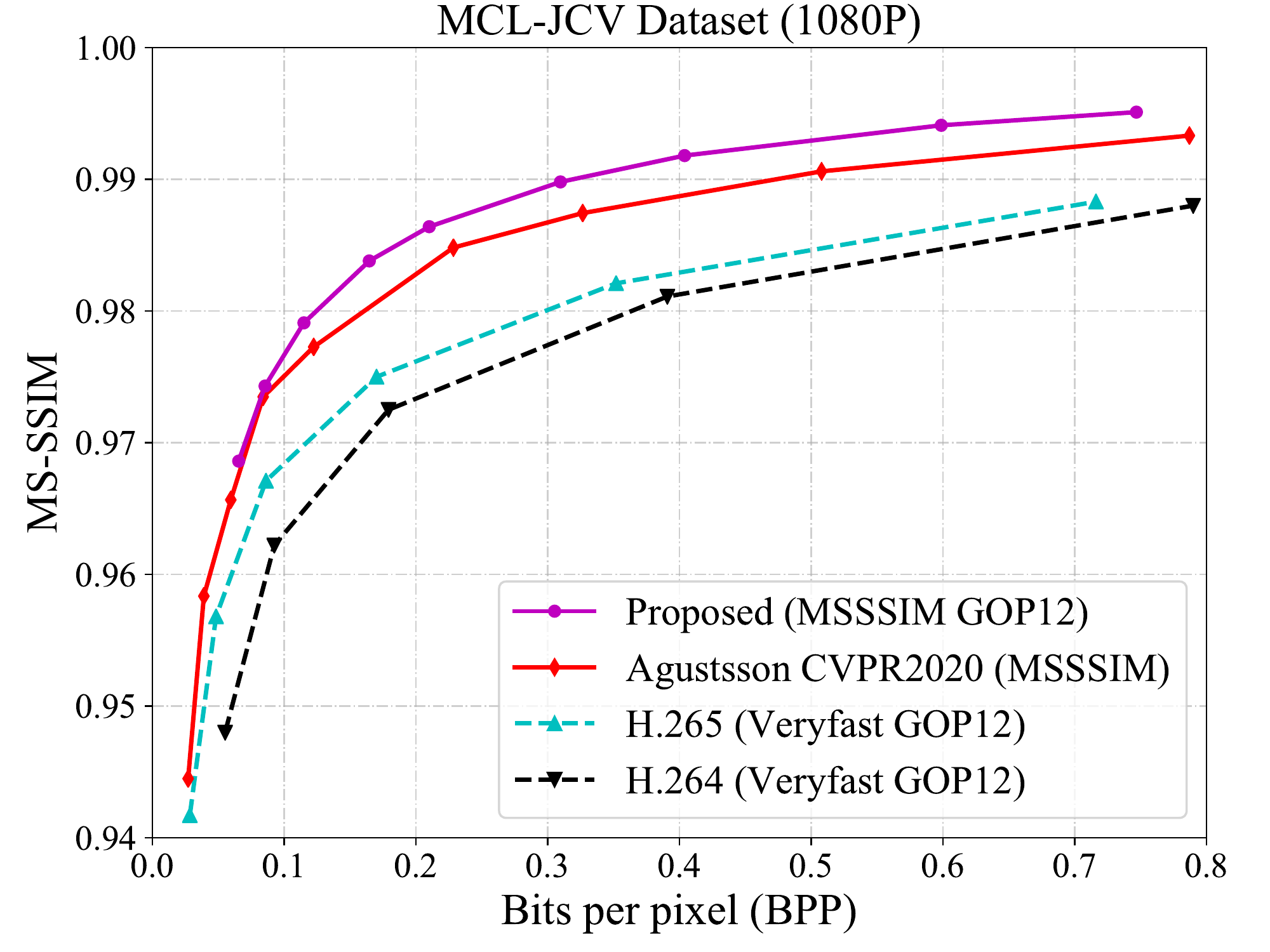}}
	\subfigure{\includegraphics[scale=0.25]{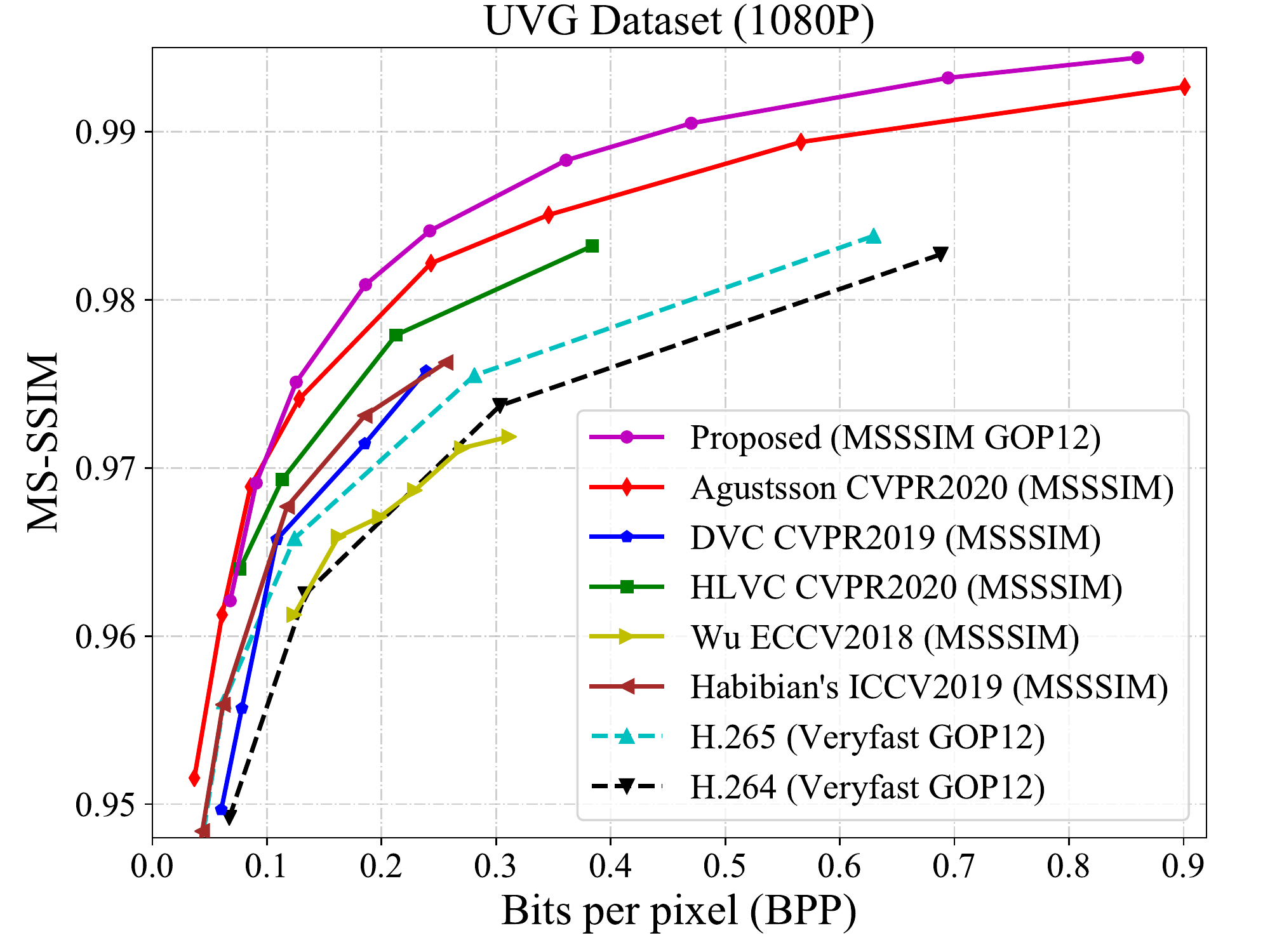}}
	\subfigure{\includegraphics[scale=0.25]{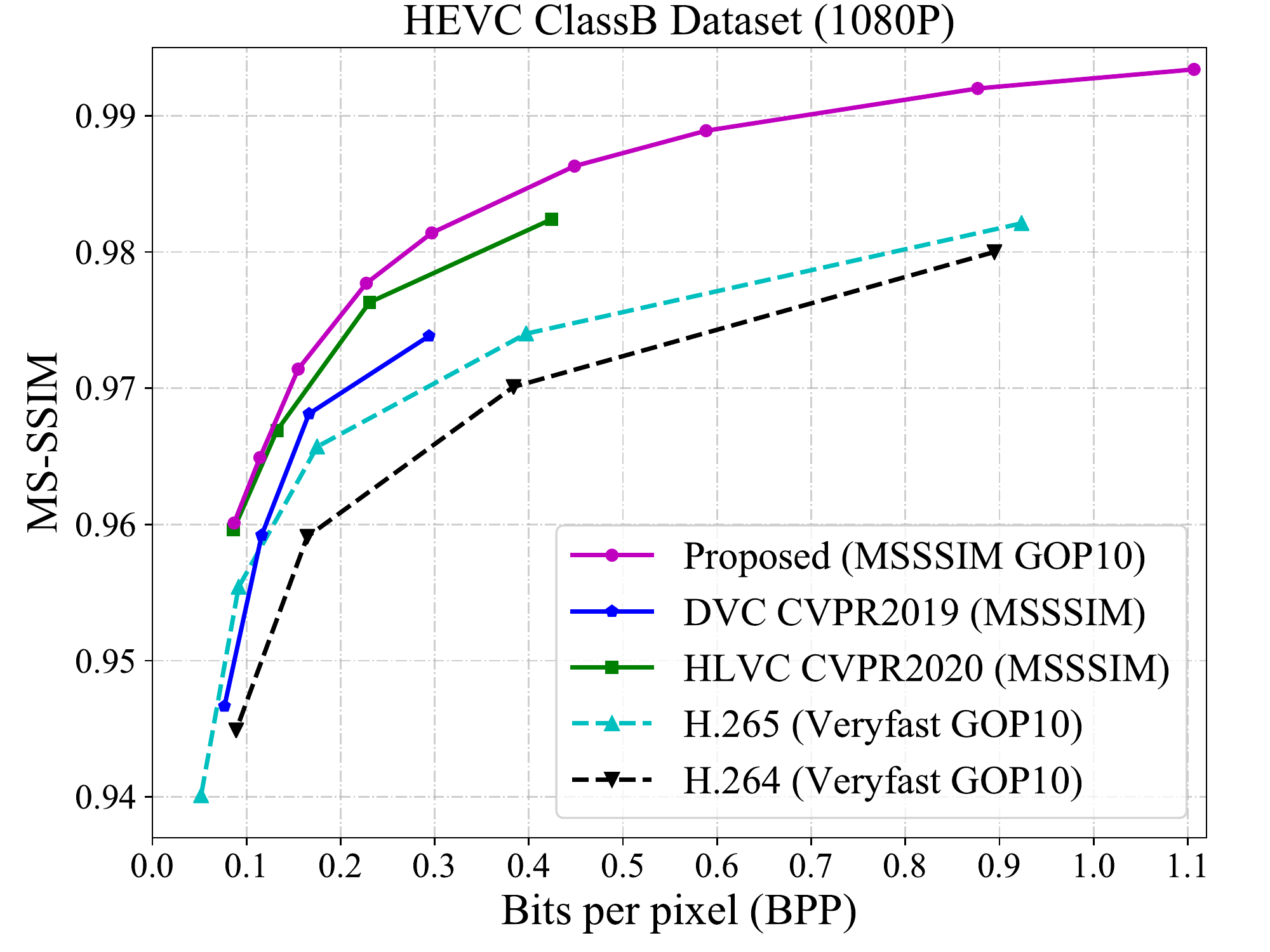}}
	\subfigure{\includegraphics[scale=0.25]{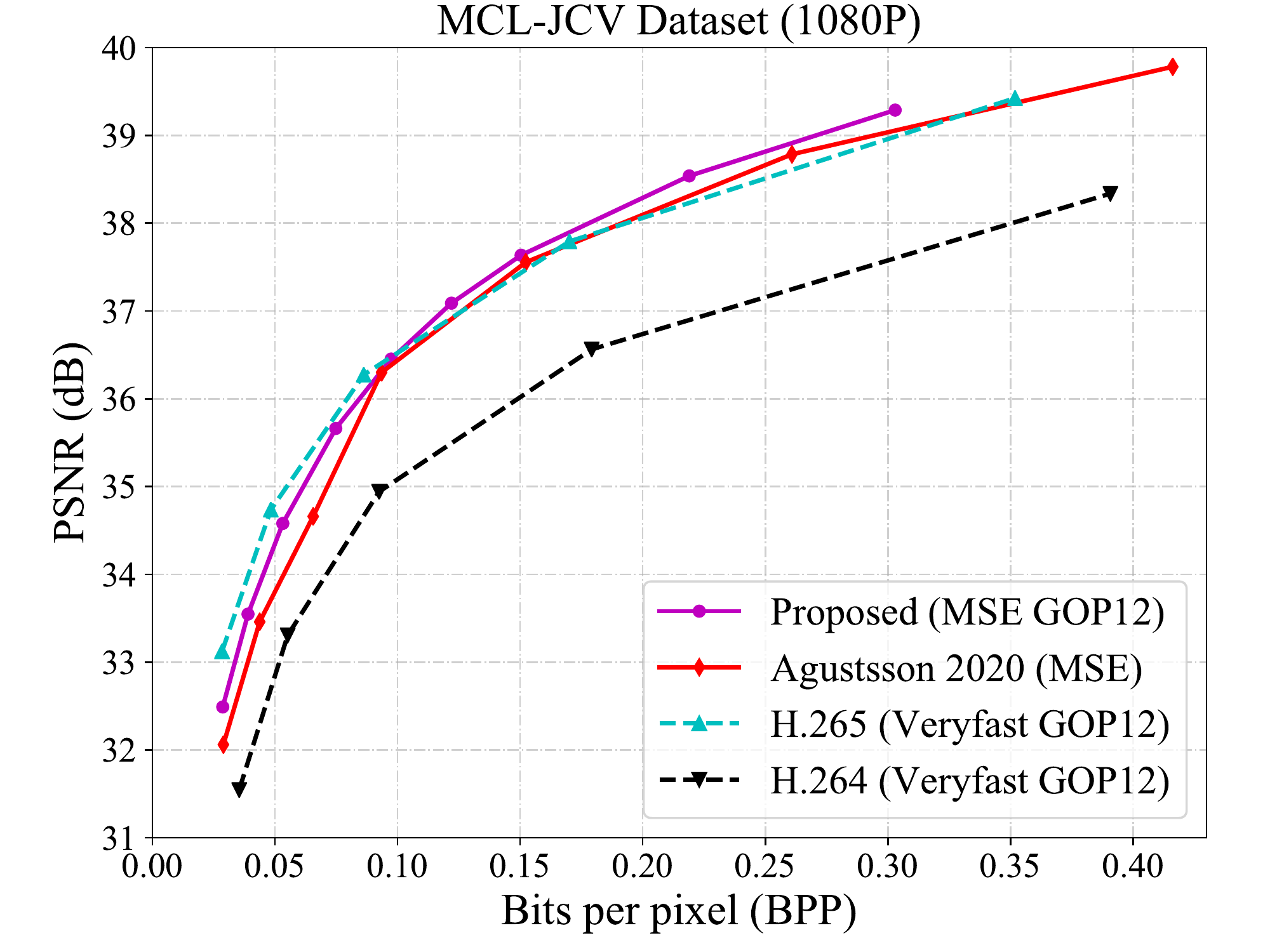}}
	\subfigure{\includegraphics[scale=0.25]{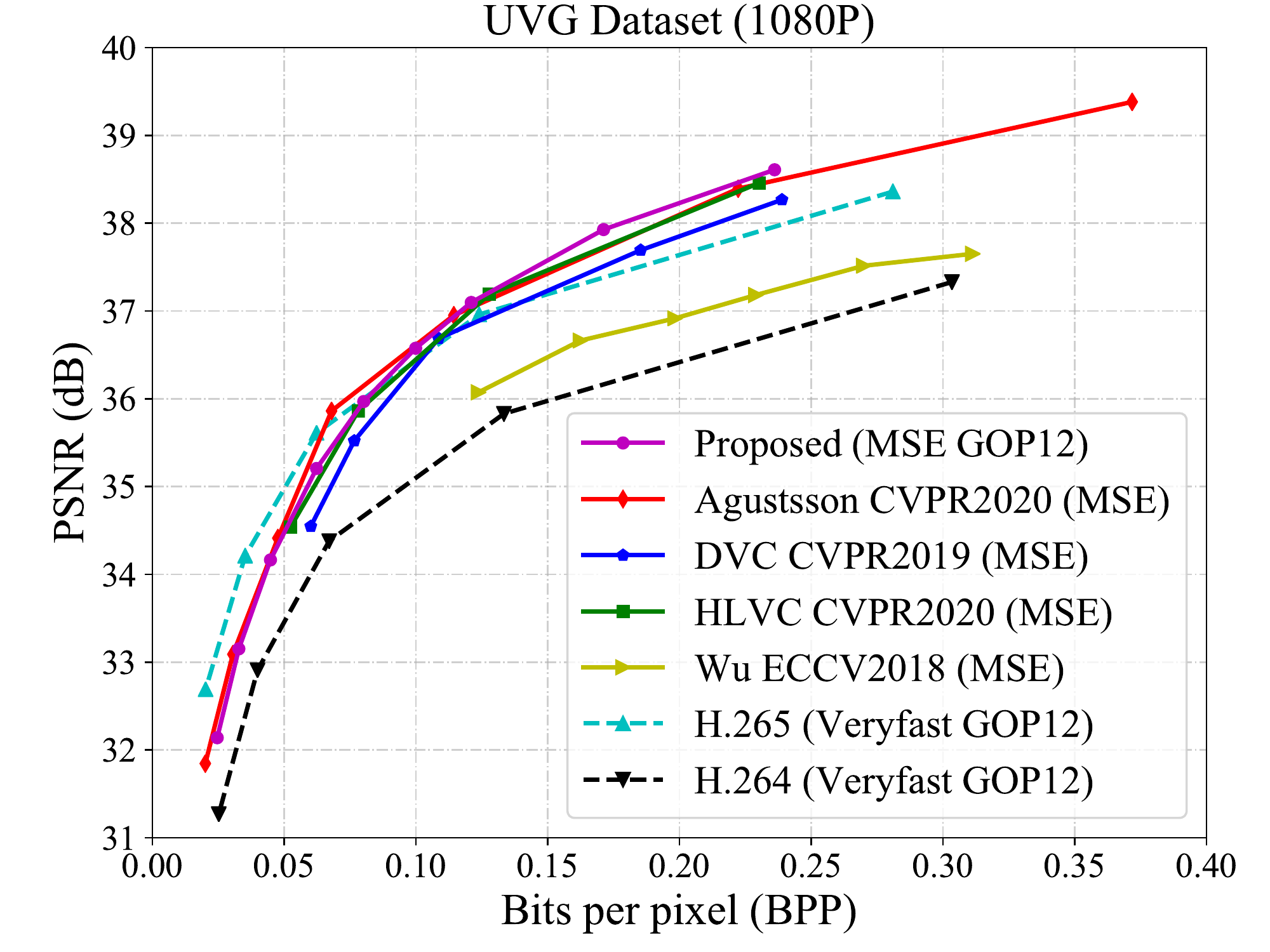}}
	\subfigure{\includegraphics[scale=0.25]{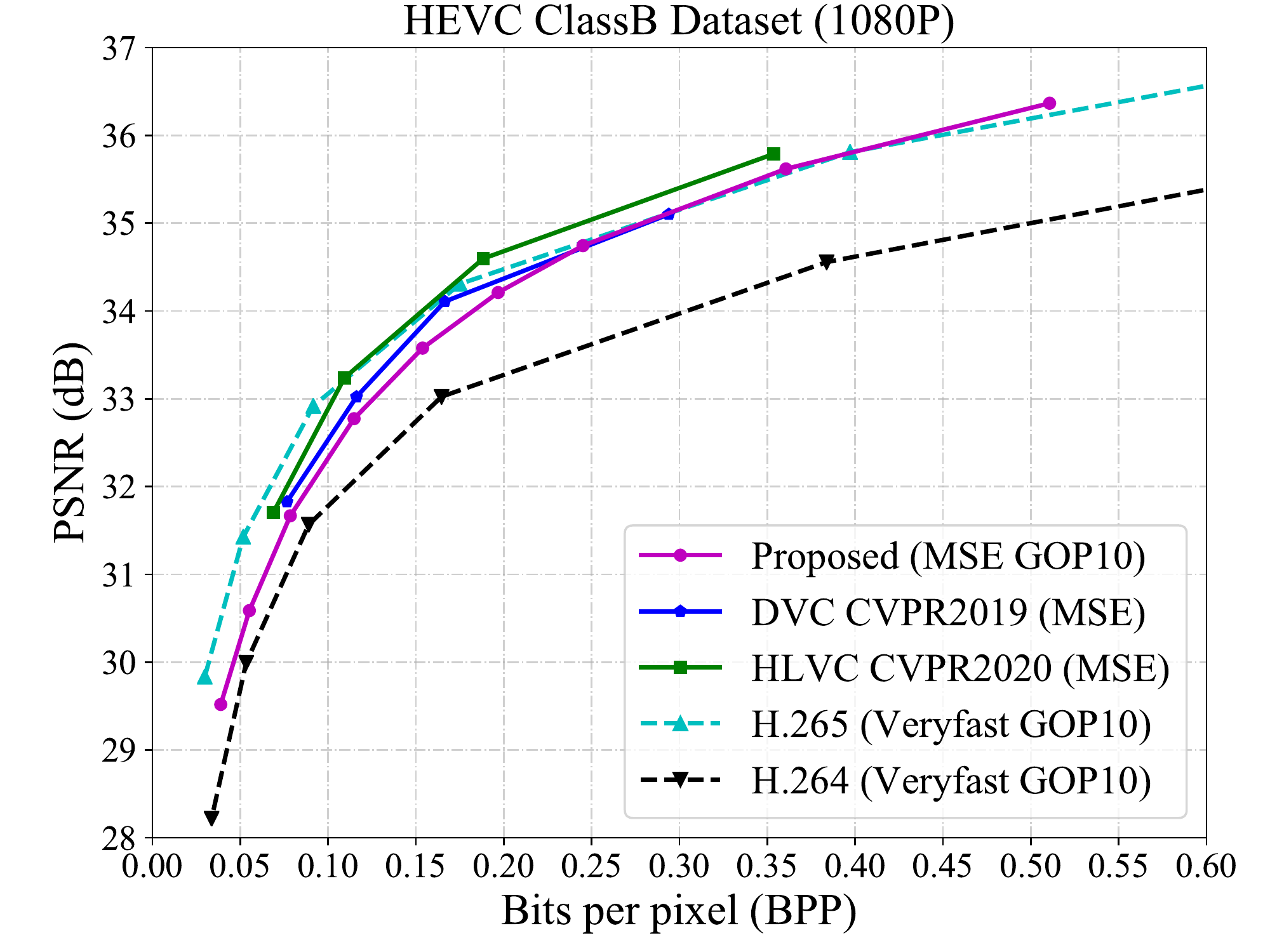}}
	\caption{Rate–distortion performance on the MCL-JCV dataset~\cite{mcl}, UVG dataset~\cite{uvg}, and HEVC Class B dataset~\cite{hevc}. Comparisons between the proposed model with the Agustsson's~\cite{google2020}, Habibian's~\cite{3D2019}, DVC~\cite{dvc}, HLVC~\cite{hlvc}, Wu's~\cite{wu2018video}, H.264~\cite{avc} and H.265~\cite{hevc}. Compared with PSNR, MS-SSIM is generally more related to perceptual quality, especially at low bit rates.}
	\label{fig:RD}
	\vspace{-0.2cm}
\end{figure*}

\subsection{Implementation Details}
{\textbf{Details For Datasets.} $\space$}
The unified auto-encoder model was trained as I-frame compression on 30K color PNG images with high resolution, scraped from Flickr~\cite{flickr}, and CLIC training dataset~\cite{clic}. The Vimeo-90k dataset~\cite{vimeo90k} is a widely used dataset for low-level vision tasks and has been applied to learned video compression tasks recently, so we used it for the P-frame training. To evaluate the compression performance and compare it with other methods, we employed three datasets with the resolution of 1080P (1920x1080).
HEVC common test sequences~\cite{hevc} are the most popular test sequences for evaluating video compression performance, in which the contents are diversified and challenging. For a regular comparison with~\cite{dvc,hlvc}, We used Class B (1080P) in experiments. 
Ultra Video Group (UVG) dataset~\cite{uvg} is composed of 16 versatile test video sequences captured at 120 fps, in which the motion between adjacent frames is small. 
MCL-JVC dataset~\cite{mcl} consists of 30 movie clips with a variety of scenarios, which contains less noise and has been used for video quality assessment~\cite{avc}.

{\textbf{Details For Training.} $\space$}
In the I-frame training of the unified auto-encoder model, the network was optimized with a batch size of 8 and a patch size of $256\times256$ randomly extracted from the training dataset. Two quality metrics (\emph{i.e.}, MSE and MS-SSIM) were used in the unified auto-encoder model.
In particular, Adam~\cite{adam} was used with multistage learning rates ($\{1e-4, 5e-5, 1e-5, 5e-6, 1e-6\}$) that changed with iteration boundaries ($\{1600000, 2100000, 2300000, 2400000, 2500000\}$). The value of Lagrange multiplier $\lambda$ was chosen from $\{50, 105, 160, 300, 480, 710, 1000, 1780, 2915\}$ for MSE and $\{3, 5, 8, 14, 20, 35, 52, 98, 145\}$ for MS-SSIM. 

In the P-frame training period, we extracted 5 interval fragments from each video sequence of Vimeo-90k, with 7 consecutive frames in each fragment. To search for different motions, the first frame was fixed as the reference frame and the current frame was randomly selected from six subsequent frames. The spatiotemporal entropy model was optimized with a batch size of 16 and a patch size of $256\times512$ randomly extracted 2 frames from the 7-frame fragment. Adam was also used with multistage learning rates ($\{1e-4, 5e-5, 1e-5, 5e-6, 1e-6\}$) that changed with iteration boundaries ($\{1200000, 1600000, 1800000, 1900000, 1950000\}$). The training was conducted in a distributed way with 4 Tesla V100, taking 42 hours for the unified auto-encoder model and 32 hours for the spatiotemporal entropy model. Details of the spatiotemporal entropy model are shown in Table.~\ref{table:IPC2}.

{\textbf{Details For Inference.} $\space$}
We evaluated the compression performance on three video test datasets. 
To evaluate the RD performance, the rate was measured by bits per pixel (BPP), and the quality was measured by either PSNR or MS-SSIM. 
The RD curves are drawn to demonstrate their coding efficiency in Fig~\ref{fig:RD}. 
To compare the compression performance fairly, we followed the settings in~\cite{wu2018video,dvc,hlvc,google2020}. 
100 frames were compressed with a GOP size of 10 for HEVC Class B sequences, and all frames were compressed with a GOP size of 12 for the UVG dataset and MCL-JCV dataset. 
\textbf{\emph{FFmpeg~\cite{ffmpeg} was used to evaluate the performance of H. 264 and H. 265 in ‘very fast’ mode with exact settings exhibited in Appendix.}}

\subsection{Performance Comparison}
{\textbf{RD Performance.} $\space$}
Fig.~\ref{fig:RD} demonstrates the RD performance of different methods on three datasets, using PSNR and MS-SSIM metrics respectively. The proposed model with 9 variable rates is compared with well-known compression standards (\emph{e.g.}, H. 264~\cite{avc} and H. 265~\cite{hevc}) and learned video compression methods (\emph{e.g.}, Wu's~\cite{wu2018video}, DVC~\cite{dvc}, Habibian's~\cite{3D2019}, HLVC~\cite{hlvc}, Agustsson's~\cite{google2020}). RD curves of Wu's, DVC, and HLVC are from the release in their GitHub, and RD curves of Agustsson's and Habibian's are obtained from the authors. Methods of Wu's and HLVC are designed for B-frame video compression, which is generally better than the P-frame compression because more reference frames are used. Fig.~\ref{fig:RD} illustrates that the proposed model performs significantly better than other methods under the metric of MS-SSIM, saving 11.10\%, 24.52\%, 31.43\%, 37.02\%, 55.86\% bits (bits-saving is all measured by BDBR~\cite{bdbr}) compared with Agustsson's, HLVC, Habibian's, DVC and Wu's on the UVG dataset, respectively. In terms of PSNR, the RD performance of the proposed model is better than that of the other learned methods on UVG/MCL-JCV datasets and is competitive to that of H.265. The lower-right figure in Fig.~\ref{fig:RD} reveals that the RD performance of the proposed model is slightly worse than that of other learned methods on HEVC Class B dataset in terms of PSNR.

The performance of the proposed model under the metric of MS-SSIM is better than that of PSNR, which might be explained that the temporal compaction is conducted in the latent space instead of the pixel space, and more attentions are paid to structure information rather than pixel information. Meanwhile, because HEVC Class B dataset has more background noise than the other datasets, the proposed model could not recover the noise to meet the metric PSNR.  Although there is some performance deficiency on the HEVC Class B in terms of PSNR, the proposed model is just a feasible attempt to the MFVC framework, which illustrates that the MFVC framework has more potential for optimization.

\begin{table}[ht]
	\begin{center}
		\renewcommand\arraystretch{1.2}
		\setlength{\abovecaptionskip}{0pt}%
		\setlength{\belowcaptionskip}{0pt}%
		\scalebox{0.57}{
		\begin{tabular}{|c|c|c|c|c|c|c|c|c|}
				\hline
				& \multicolumn{2}{c|}{Encoder}  & \multicolumn{2}{c|}{Decoder}  & UVG (BDBR) & Class B (BDBR)  \\
				\hline
				Model & FLOPs & Latency & FLOPs & Latency & MS-SSIM & MS-SSIM\\
				\hline
				DVC~\cite{dvc} & 3074G & 426ms & 1434G & 189ms & -25.24\% & -35.35\%\\
				HLVC~\cite{hlvc} & 2831G & 376ms & 1246G & 184ms & -49.00\% & -56.87\%\\
				Proposed w/o SPM & 613G & 104ms& 1479G & 121ms & -55.70\% & -61.07\% \\
			    \hline
				Proposed & 643G & 114ms & 1509G & \textcolor{red}{23s} & -60.29\% & -62.73\% \\
				\hline
		\end{tabular}}
	\end{center}
	\caption{Floating-point performance on a Tesla V100 with an input p-frame of 1080P (\emph{i.e.}, resolution: $1920\times1080\times3$) except for the decoding of the proposed. Due to the serial characteristic of the PixelCNN, the decoding of the proposed is conducted on an 8-core CPU. BDBR is measured by setting H.264 as the anchor. Note that the proposed method is in redundancy design, and the channels are more than 256. The channels of DVC and HLVC are 128 or less.}
	\label{table:fpp}
	\vspace{-0cm}
\end{table}

{\textbf{Floating Point Performance.} $\space$}
We compared the floating-point performance between the proposed methods and the open-source methods (i.e, DVC, and HLVC-layer2-P). 
We tested the first P-frame compression for all three methods and the values are summarized in Table.~\ref{table:fpp}.
The results reveal that the proposed w/o SPM saves 3-4 times of inference time in the encoding period and saves 30\% inference time in the decoding period.
Meanwhile, comparing the RD performance on the UVG dataset, the proposed w/o SPM saves 30.46\% (DVC) and 6.7\% (HLVC) bits in terms of MS-SSIM.
Due to the redundancy design, the complexity of the network can be reduced to further speed up the inference in the future.

\subsection{Ablation Study}\label{sec:44}
\vspace{-0.1cm}

\begin{figure}[h]
	\centering
	\includegraphics[scale=0.4]{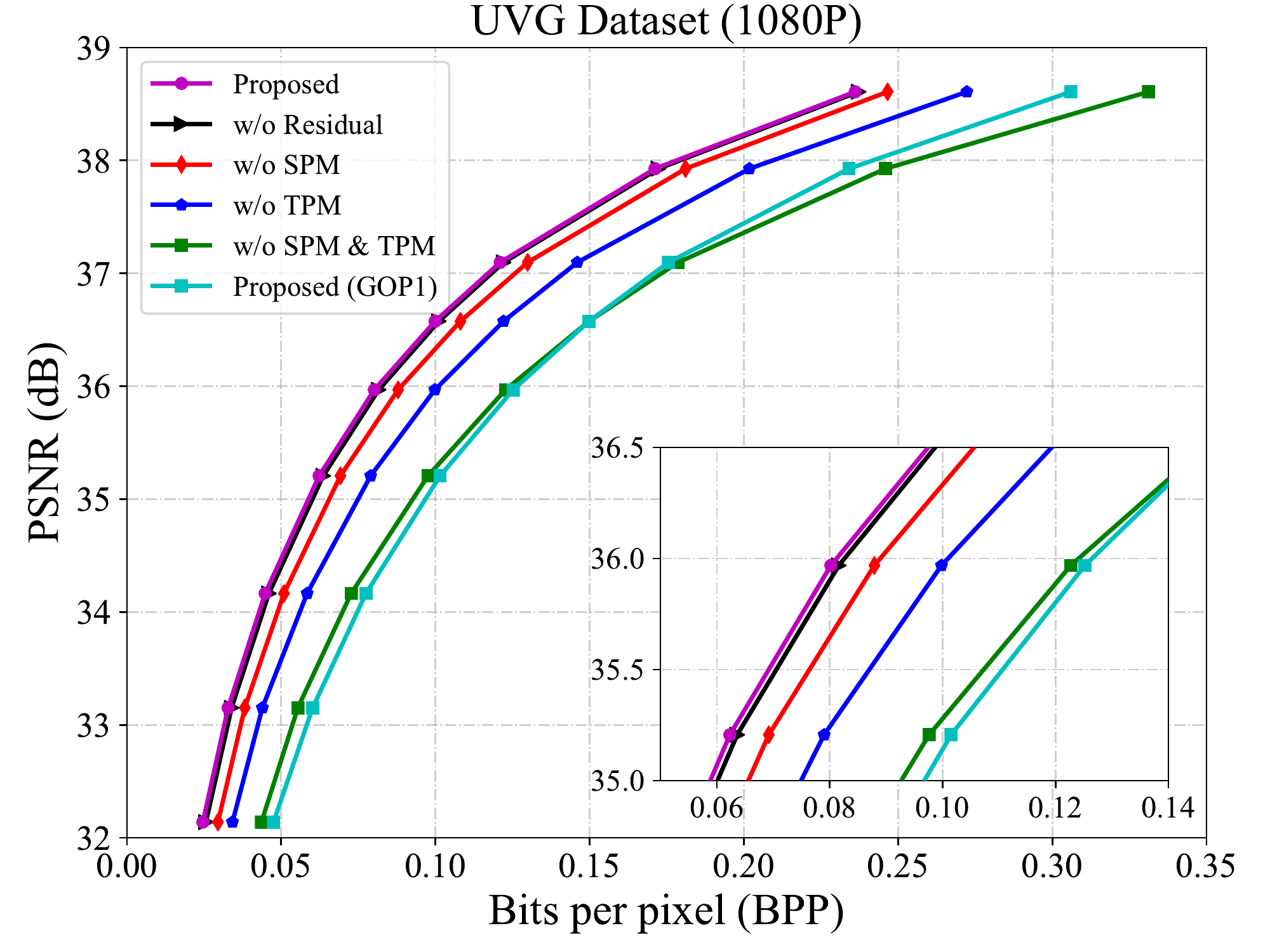}
	\caption{Ablation study on the UVG dataset. All variants are optimized by MSE and the test GOP size is set as 12 except for the {\it{Proposed (GOP1)}}. {\it{w/o Residual}}: the proposed method removes the residual function and $\hat{y}_t$ is directly used for the probability prediction in the spatiotemporal entropy model.}
	\label{fig:ablation}
	\vspace{-0cm}
\end{figure}

To verify the effectiveness of each modular component in the proposed model, a set of ablation experiments were conducted on the UVG dataset and RD curves are drawn in Fig.~\ref{fig:ablation}. 
By setting the I-frame compression (GOP size of 1) as the anchor, the proposed model saves $37.71\%$ bits while maintaining the same reconstructive quality. 
If $\hat{y}_t$ is directly used for the probability prediction, the proposed model ({\it{w/o Residual}}) saves $36.43\%$ bits. 
If the SPM and the TPM are not used respectively, the proposed model saves $30.76\%$ bits (w/o SPM) and $21.32\%$ bits (w/o TPM) accordingly. 
If SPM and TPM are both removed, the proposed model degrades to the I-frame compression and only saves $2.56\%$ bits (w/o SPM \& TPM). 
This reveals that SPM and TPM both contribute to accurate probability prediction, leading to a bits-saving of $6.95\%$ (SPM), $16.39\%$ (TPM), and $35.15\%$ (Both) on the UVG dataset.

\begin{figure}[t]
	\centering
	\includegraphics[scale=0.65]{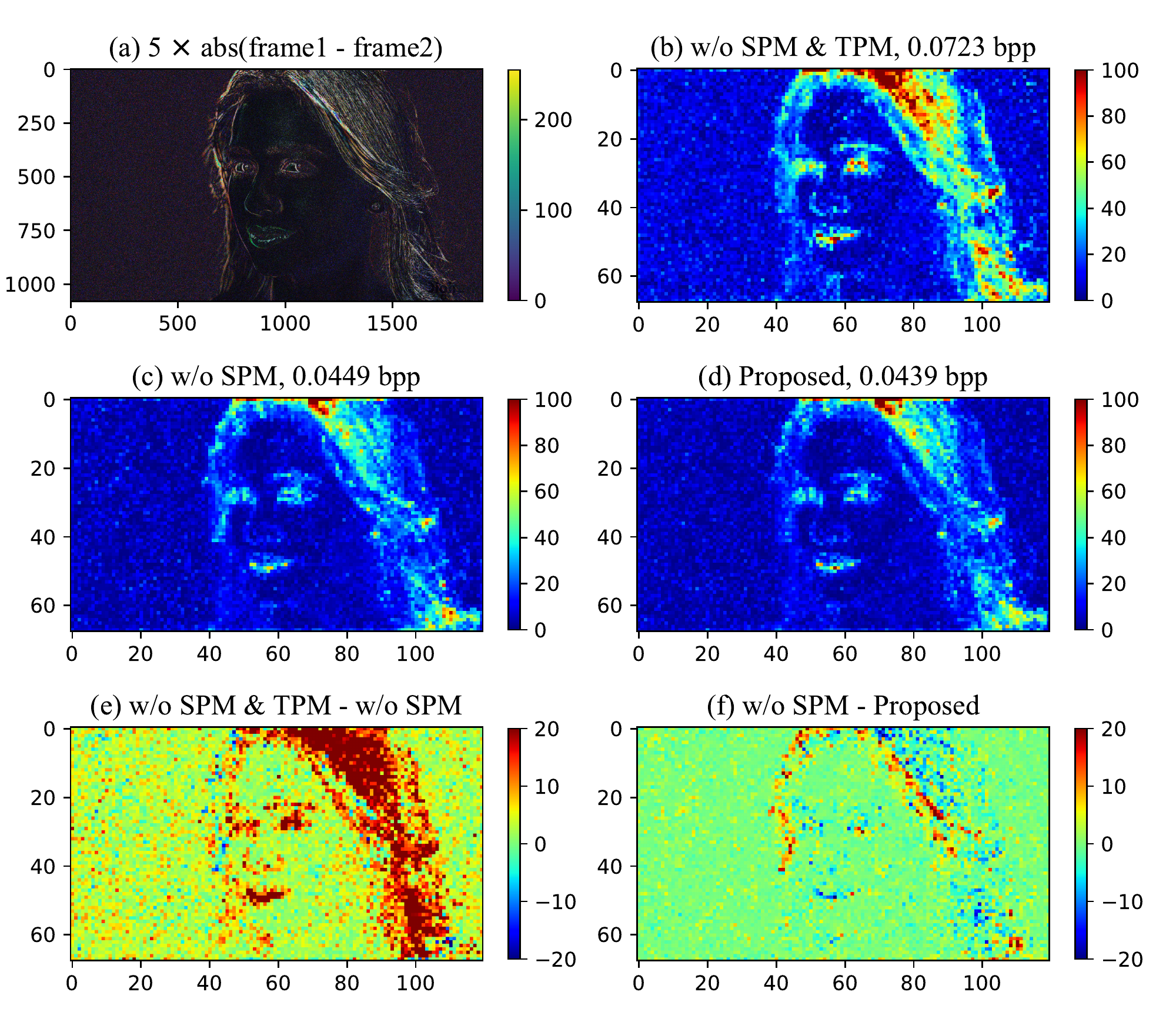}
	\caption{Visualizations on the \emph{Beauty} sequences. (a) is the differential results between the adjacent frames, showing the motion (\emph{e.g.}, fluttering hair) in the pixel level. (b), (c), and (d) are heatmaps of entropy values that represent bits per pixel. (e) and (f) are the differential values of (b)-(c) and (c)-(d), respectively.}
	\label{fig:visual_abla}
	\vspace{-0.2cm}
\end{figure}

Additionally, we visualized the entropy values to further distinguish the difference between the variants, and visualizations are displayed in Fig.~\ref{fig:visual_abla}. The motion information between adjacent frames is shown in Fig.~\ref{fig:visual_abla} (a). And the entropy values of the proposed w/o SPM \& TPM,  w/o SPM, and the proposed are displayed in Fig.~\ref{fig:visual_abla} (b), (c), and (d), respectively. 
Through observing Fig.~\ref{fig:visual_abla} (b), (c), and their differential results Fig.~\ref{fig:visual_abla} (e), the TPM can reduce more entropy in motion area (mainly on fluttering hair) due to exploiting the temporal redundancy. The differential results (Fig.~\ref{fig:visual_abla} (f)) between Fig.~\ref{fig:visual_abla} (d) and (c) demonstrate the SPM can save some bits on the outline and background by referring to neighborhood points in front of them.


\section{Conclusion}
In this paper, we mainly discuss an open challenge that whether we could achieve competitive performance with dominant learned video compression methods in a simple framework. Firstly, we compare the dominant framework of learned based methods and that of traditional video coding schemes. It is shown that the current learned one suffers from its complex structure and error propagation problems. Secondly, we discuss some efforts, \emph{e.g.}, error propagation aware training, and implicit flow network, to solve these inherent issues. Although the methods have brought some improvement, little attention has been paid to the framework itself. Finally, a learned image compression style framework is proposed in this paper, called motion-free video compression (MFVC). It utilizes a variable-rate auto-encoder and a spatiotemporal entropy model to outperform SOTA performance under the metric of MS-SSIM. Last but not the least, the MFVC indicates that we may attempt more advanced sequence technologies, \emph{e.g.}, ConvLSTMs~\cite{convlstm}, 3D Convolution, explicit/implicit optical flow, and transformers, on entropy model rather than on raw pixels. \textbf{\emph{More experiments results are present in the Appendix.}}


{\small
	\bibliographystyle{ref_fullname}
	\bibliography{reference}
}

\newpage
\appendix
\section{Appendix}
\subsection{Distortion of Motion Prediction}
In the main body of the paper, we mentioned that dominant learned video compression methods suffer from three inherent problems caused by the motion-based video compression framework, and two problems (\emph{i.e.}, the complexity and error propagation problem) are already present in section 3.1. In this section, we will explain the second problem that the inaccuracy of a predictive frame may result in residual errors and then bring more difficulty to the residual compression in the motion-based video compression method.
An accurate optical flow network (\emph{i.e.}, PWCnet~\cite{pwc}) was used as the motion estimation module in motion-based video compression framework. The compression process is evaluated on three different video sequences, \emph{e.g.}, \emph{Beauty} and \emph{YachtRide} in UVG, and \emph{BasketballDrive} in HEVC Class B. The qualitative results are visualized in Fig~\ref{fig:BT}, Fig~\ref{fig:YR} and Fig~\ref{fig:BD}.

In these figures, (a) is the previous reconstructive frame and (b) is the current frame. (c) is the optical flow between (a) and (b). (d) is the predictive frame after motion compensating (\emph{i.e.}, warped from the previous reconstructive frame with the reconstructive optical flow). (e) is the differential results (5 times higher for visualization) between (a) and (b), showing the motion information between adjacent frames. (f) is the differential results (5 times higher for visualization) between (d) and (b), and (f) is the residual image for residual compression in motion-based video compression. 
The ellipse regions in (d) are distorted regions caused by inaccurate motion prediction. The corresponding regions in (e) and (f) reveal that the residual image to be compressed will have higher temporal redundancy than the direct differential results between the previous reconstructive frame and the current frame. This phenomenon proves that motion prediction may decrease the accuracy of a predictive frame. 

\subsection{Details for Error Propagation}

\begin{figure}[h]
	\centering
	\includegraphics[scale=0.4]{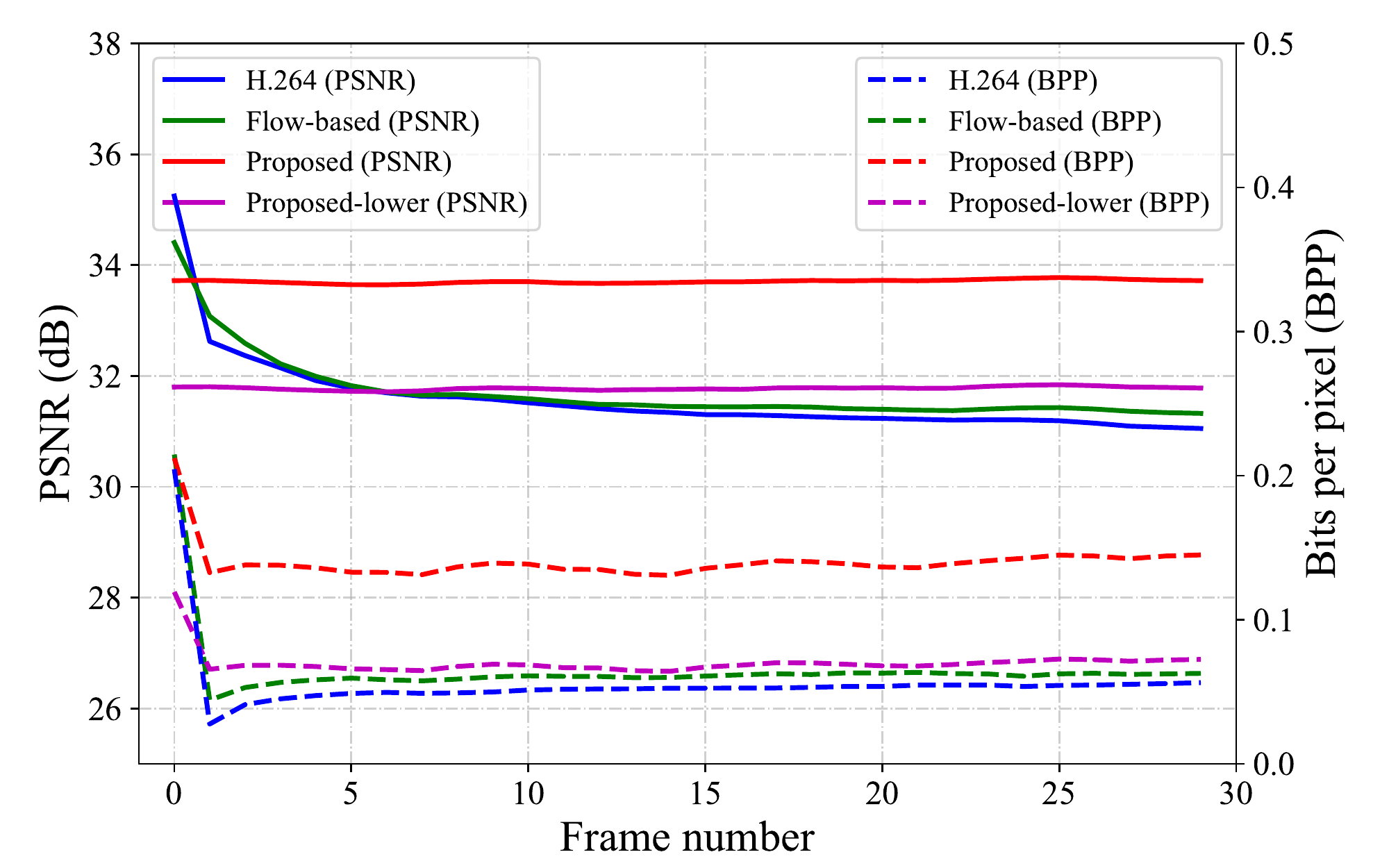}
	\caption{PSNR values for multi-frames compression on HEVC Class B dataset. H. 264 is conducted at a fixed compression rate (i.e., quantization parameter of 27). Flow-based method is a typical learned video compression with optical flow and wraping operation in a single rate model. These two methods are both based on the motion-based framework. Note that the BPP of the 0th frame on the H.265 curve is four times smaller than the actual value for intuitive visualization (from 0.82 to 0.21).}
	\vspace{-0.2cm}
	\label{fig:error}
\end{figure}

\begin{table*}[t]
	\begin{center}
		\renewcommand\arraystretch{1}
		\setlength{\abovecaptionskip}{0pt}%
		\setlength{\belowcaptionskip}{0pt}%
		\scalebox{1}{
		\begin{tabular}{|c|p{1.5cm}<{\centering}|p{1.5cm}<{\centering}|p{1.5cm}<{\centering}<{\centering}|p{1.5cm}<{\centering}|p{1.5cm}<{\centering}|p{1.5cm}<{\centering}|}
				\hline
				&\multicolumn{2}{c|}{MCL-JCV~\cite{mcl}} & \multicolumn{2}{c|}{UVG~\cite{uvg}}& \multicolumn{2}{c|}{Class B~\cite{hevc}}\\
				\cline{2-7}
				Methods & PSNR & MS-SSIM & PSNR & MS-SSIM & PSNR & MS-SSIM \\
				\hline
				Proposed & -40.75\% & -63.24\% & -47.05\% & -60.29\%&  -33.78\%& -62.73\% \\
				Proposed w/o SPM & -36.75\% & -60.42\% & -42.93\% & -57.70\%&  -31.96\%& -61.07\% \\
				H. 265~\cite{hevc} & -40.44\% & -27.86\% & -46.90\%& -36.61\%& -39.45\% & -31.67\% \\
				Wu's~\cite{wu2018video} & / & / & -16.07\%& 3.43\%& /& /\\
				DVC~\cite{dvc} & / & / & -41.17\%& -25.24\%& -37.08\%& -35.35\%\\
				HLVC~\cite{hlvc} & / & / & -46.79\%& -49.00\%& -44.50\%& -56.87\%\\
				Habibian's~\cite{3D2019} & / & / & / & -35.36\%& / & / \\
				Agustsson's~\cite{google2020} & -33.57\%& -51.51\%& -46.56\%& -53.35\%& /& /\\
				\hline
		\end{tabular}}
	\end{center}
	\caption{BD-Rate Gains of Proposed, H. 265~\cite{hevc}, Wu's~\cite{wu2018video}, DVC~\cite{dvc}, HLVC~\cite{hlvc}, Habibian's~\cite{3D2019}, and Agustsson's~\cite{google2020} against the H.264~\cite{avc}. Negative values in BDBR represent the bits saving. “/” represents that the method didn't evaluate the RD performance on the dataset in that column.}
	\vspace{-0.4cm}
	\label{table:bdbr}
\end{table*}

In the main body of the paper, figure 3 only displays the PSNR values of those three methods. In the section, more details about bits per pixel (BPP) are shown in Fig.~\ref{fig:error}. The BPP curves of H.264 and flow-based method remain stable as the frame number increases, the qualities of the reconstruction begin to drop from 35.24 dB to 31.05 dB for H.264 and from 34.41 dB to 31.324 dB for flow-based method, showing a big error propagation in a single compression rate. Additionally, the proposed motion-free video compression framework maintains the same quality level along with the frames (whether in a higher or lower bitrate).

Besides, the BPP of the P-frame is not many times lower than that of the I-frame in the proposed method due to no error propagation between the reconstructive frames. That means the proposed method does not need to sacrifice the reconstruction qualities to achieve better RD performance. Based on the proposed framework, more efforts will be paid to improve the spatiotemporal entropy model without considering the reconstruction quality in the future.
\subsection{Performance Comparison Based on BDBR}

To further compare the RD performance between different methods, the BD-rate gains measured by BDBR~\cite{bdbr} by setting H.264 as the anchor and the results are summarized in Table.~\ref{table:bdbr}. As consistent with the results of RD curves, Table.~\ref{table:bdbr} also illustrates that the proposed model performs significantly better than other methods under the metric of MS-SSIM, achieving the SOTA RD performance. In terms of PSNR, the RD performance of the proposed model is better than that of the other learned methods on UVG/MCL-JCV datasets and is competitive to that of H.265. The BDBR results also reveal that the RD performance of the proposed model is slightly worse than that of other learned methods on HEVC Class B dataset in terms of PSNR. Meanwhile, the proposed w/o SPM has a slight degradation compared with the proposed method, which is also stated in the ablation study.


\subsection{Settings of H.264 and H.265}
FFmpeg~\cite{ffmpeg} was used to evaluate the performance of H. 264 and H. 265 in {\it very fast} mode and the exact commands as following:\\\\
\textbf{H.264(\emph{very fast})}\\
ffmpeg -pix\_fmt yuv420p -s WxH -r FPS -i Name.yuv -c:v libx264 -preset very fast -tune zerolatency -crf QP -g GOP -v 1 -bf 0 Name.mkv\\\\
\textbf{H.265(\emph{very fast})}\\
ffmpeg -pix\_fmt yuv420p -s WxH -r FPS -i Name.yuv -c:v libx265 -preset very fast -tune zerolatency -x265-params 'crf=QP:keyint=GOP:verbose=1:bframes=0' Name.mkv\\\\
where 'WxH' represents resolution, 'FPS' represents frames per second, 'Name.yuv' represents the input file, QP represents quantization parameter (i.e., compression rate), GOP represents group of pitures, and 'Name.mkv' represents the out file.

\subsection{Future works}
Based on the separation of the RD performance, more efforts should be paid to promote the unified auto-encoder network for the quality of the reconstruction and improve the spatiotemporal entropy network for less compressed bitstream.
To further enhance the performance of the proposed framework, an interesting topic is to use multiple latents for the spatiotemporal entropy network, such as the ConvLSTMs~\cite{convlstm} for temporal relationships, or B-frame compression. 
To make up the gap with traditional video codecs, such as H.264 and H.265, the input of learned video compression may be consistent with the traditional codecs, exploring the training methods in YUV color space. 
To be adaptive to the different video contents, some works need to be conducted to explore finetuning the Encoder in the proposed framework.

\subsection{Subjective Comparision }
In Fig~\ref{fig:BTComp} and Fig~\ref{fig:YRComp}, we visualize the reconstructive frames taken from two different evaluation video sequences \emph{Beauty} and \emph{YachtRide} in UVG. Comparisons are conducted between the proposed method, H.264 and H.265. The experimental results verify that our proposed methods save more bits at the same reconstruction quality level.

\newpage
\begin{figure*}[h]
	\centering
	\subfigure{\includegraphics[scale=0.4]{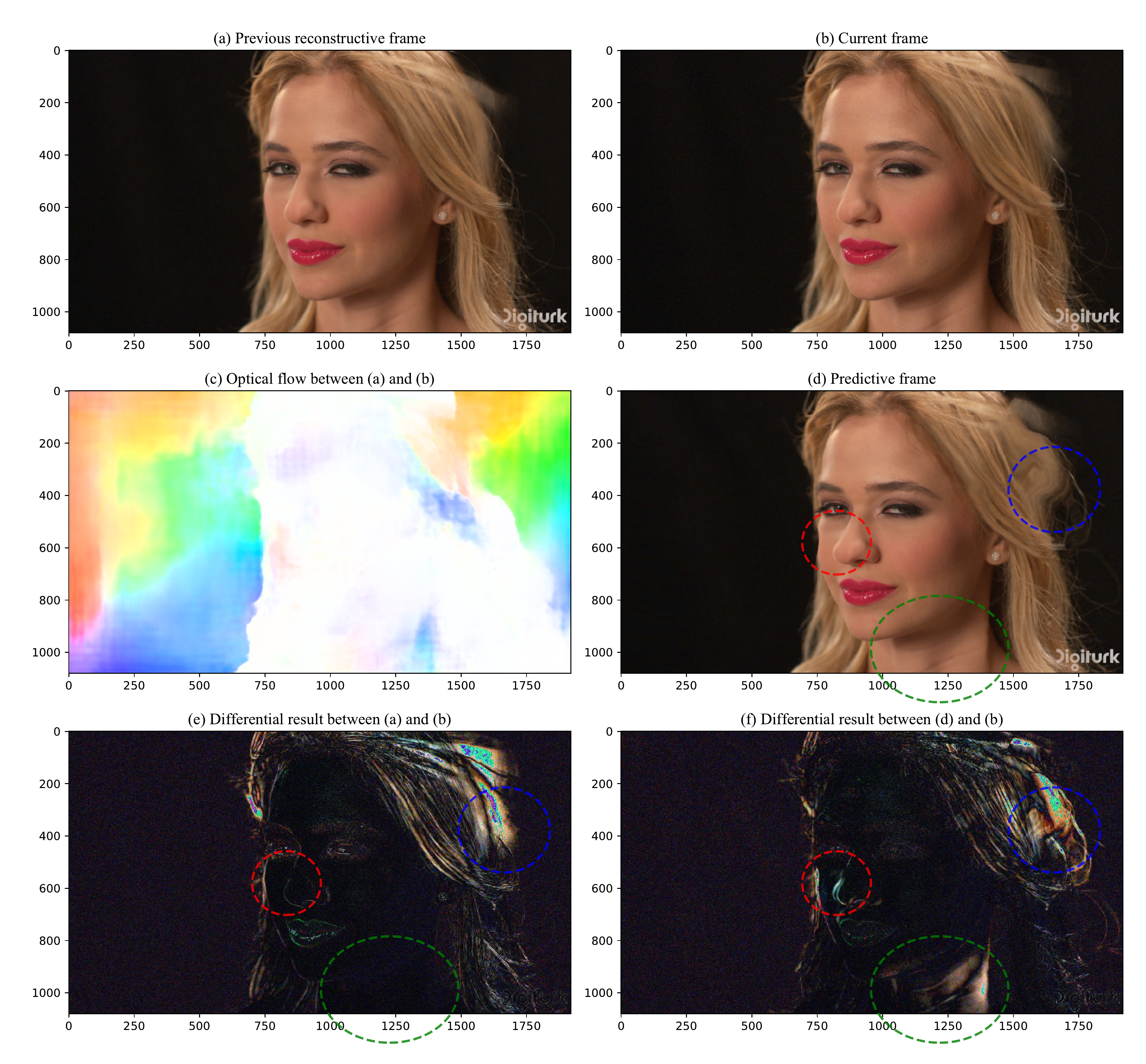}}
	\caption{Visualizations of the motion prediction process on the \emph{Beauty} sequences in UVG dataset. (a) is the previous reconstructive frame. (b) is the current frame. (c) is the optical flow between (a) and (b). (d) is the predictive frame after motion compensating (\emph{i.e.},warped from the previous reconstructive frame with the reconstructive optical flow). (e) is the differential results (5 times higher for visualization) between (a) and (b), showing the motion information between adjacent frames. (f) is the differential results (5 times higher for visualization) between (d) and (b), and (f) is the residual image for residual compression in the motion-based video compression. The ellipse regions in (d) are distorted regions caused by inaccurate motion prediction. The corresponding regions in (e) and (f) reveal that the residual image to be compressed will have higher temporal redundancy than the direct differential results between the previous reconstructive frame and the current frame.}
	\label{fig:BT}
\end{figure*}

\begin{figure*}[h]
	\centering
	\subfigure{\includegraphics[scale=0.4]{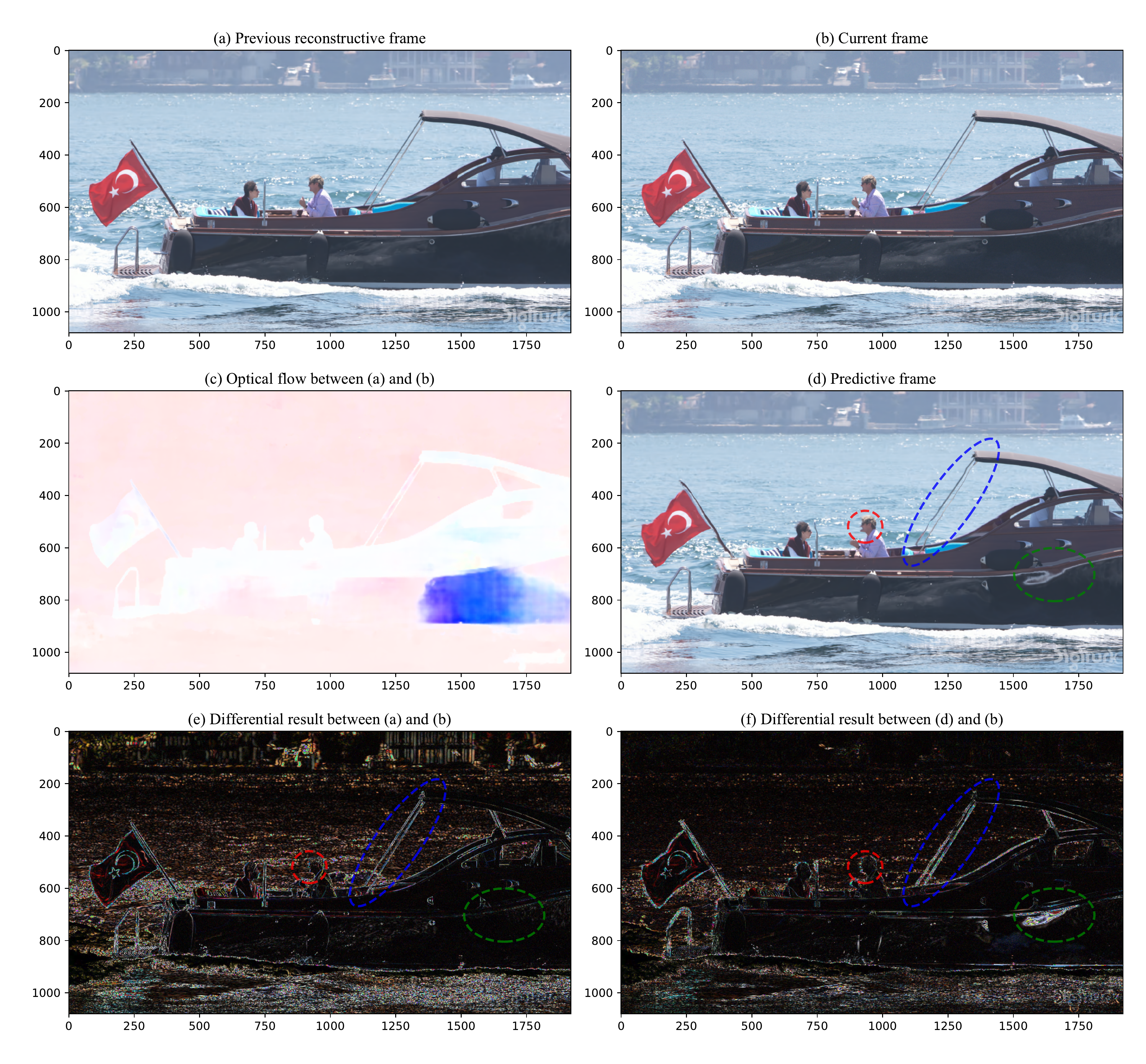}}
	\caption{Visualizations of the motion prediction process on the \emph{YachtRide} sequences in UVG dataset. (a) is the previous reconstructive frame. (b) is the current frame. (c) is the optical flow between (a) and (b). (d) is the predictive frame after motion compensating (\emph{i.e.},warped from the previous reconstructive frame with the reconstructive optical flow). (e) is the differential results (5 times higher for visualization) between (a) and (b), showing the motion information between adjacent frames. (f) is the differential results (5 times higher for visualization) between (d) and (b), and (f) is the residual image for residual compression in the motion-based video compression. The ellipse regions in (d) are distorted regions caused by inaccurate motion prediction. The corresponding regions in (e) and (f) reveal that the residual image to be compressed will have higher temporal redundancy than the direct differential results between the previous reconstructive frame and the current frame.}
	\label{fig:YR}
\end{figure*}

\begin{figure*}[h]
	\centering
	\subfigure{\includegraphics[scale=0.4]{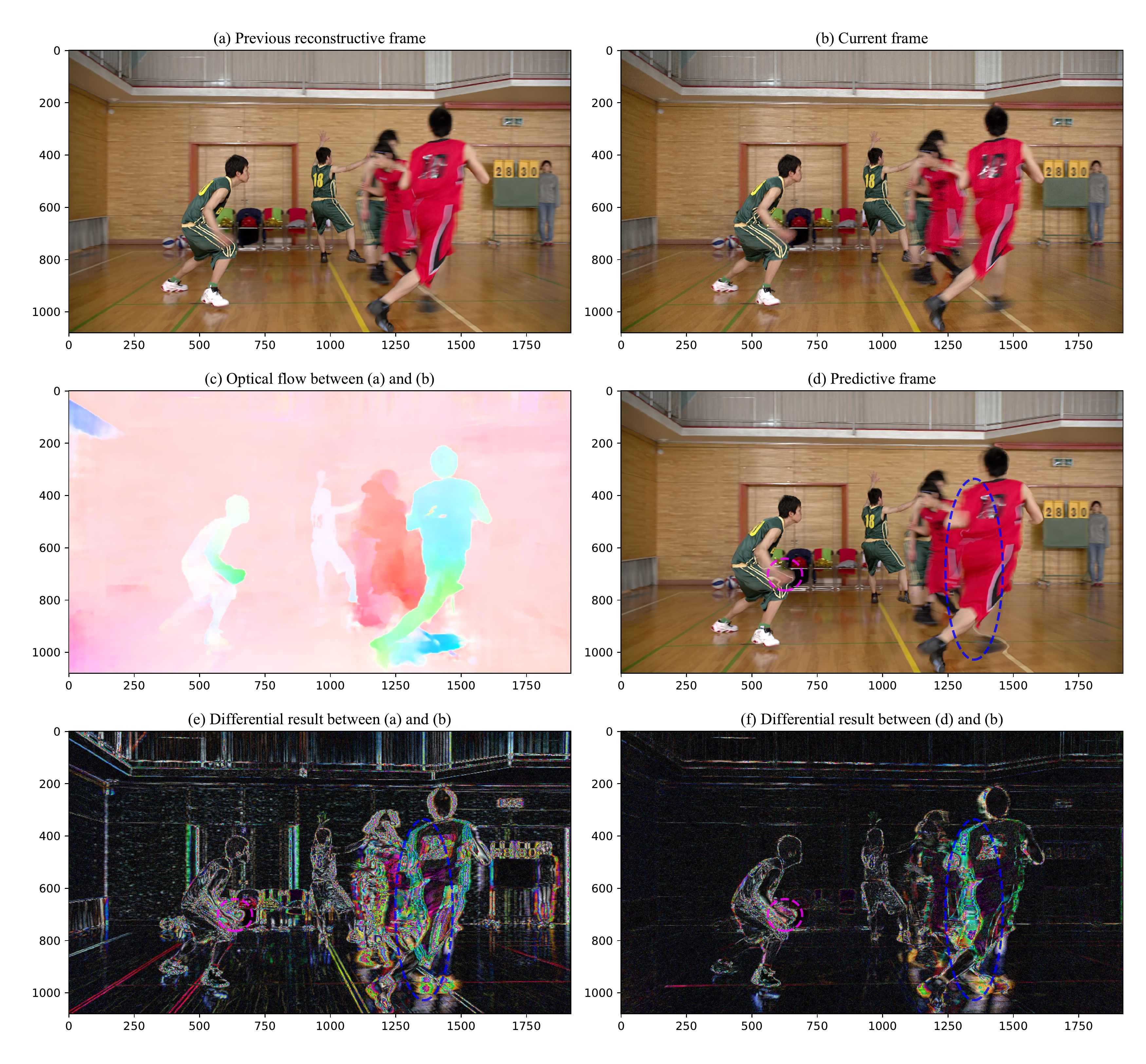}}
	\caption{Visualizations of the motion prediction process on the \emph{BasketballDrive} sequences in HEVC Class B dataset. (a) is the previous reconstructive frame. (b) is the current frame. (c) is the optical flow between (a) and (b). (d) is the predictive frame after motion compensating (\emph{i.e.},warped from the previous reconstructive frame with the reconstructive optical flow). (e) is the differential results (5 times higher for visualization) between (a) and (b), showing the motion information between adjacent frames. (f) is the differential results (5 times higher for visualization) between (d) and (b), and (f) is the residual image for residual compression in the motion-based video compression. The ellipse regions in (d) are distorted regions caused by inaccurate motion prediction. The corresponding regions in (e) and (f) reveal that the residual image to be compressed will have higher temporal redundancy than the direct differential results between the previous reconstructive frame and the current frame.}
	\label{fig:BD}
\end{figure*}

 \begin{figure*}[h]
 	\centering
 	\subfigure[H.264, 0.1411 bpp, 34.00 dB, 0.9079]{\includegraphics[scale=0.10]{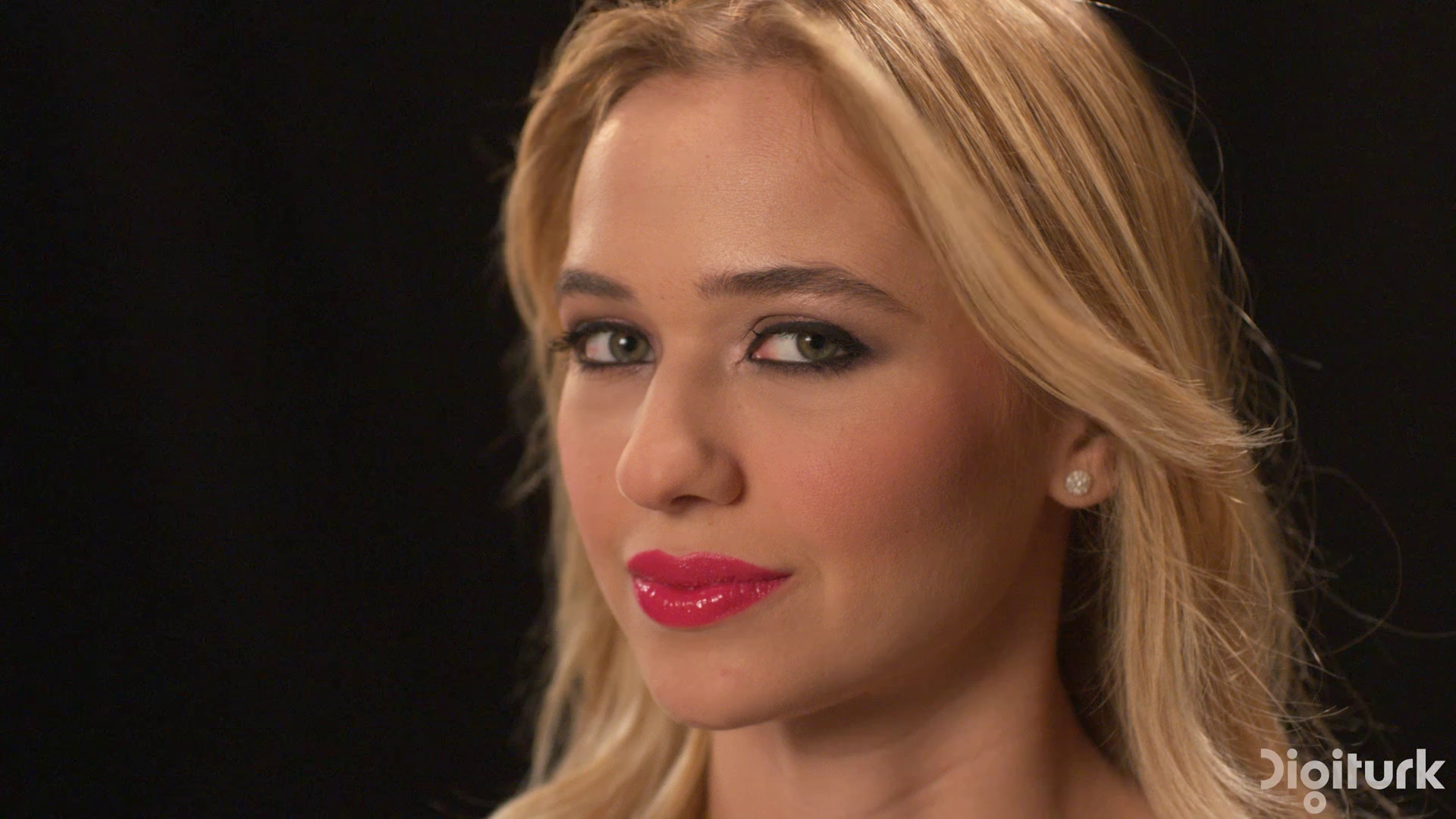}}
 	\subfigure[H.265, 0.1133 bpp, 34.16 dB, 0.9078]{\includegraphics[scale=0.10]{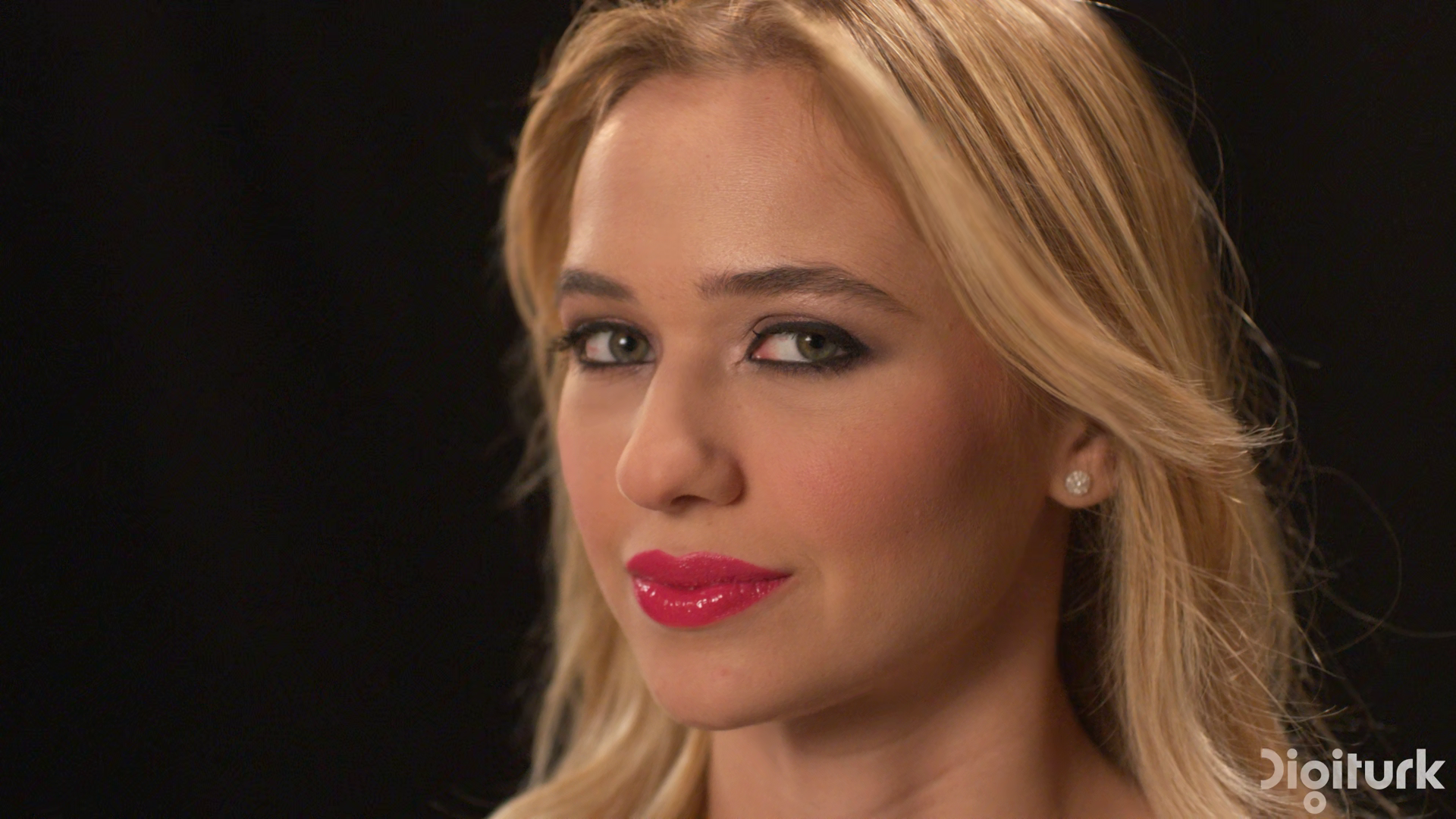}}
 	\subfigure[Proposed (MSE), 0.0609 bpp, 34.0251 dB, 0.9072]{\includegraphics[scale=0.10]{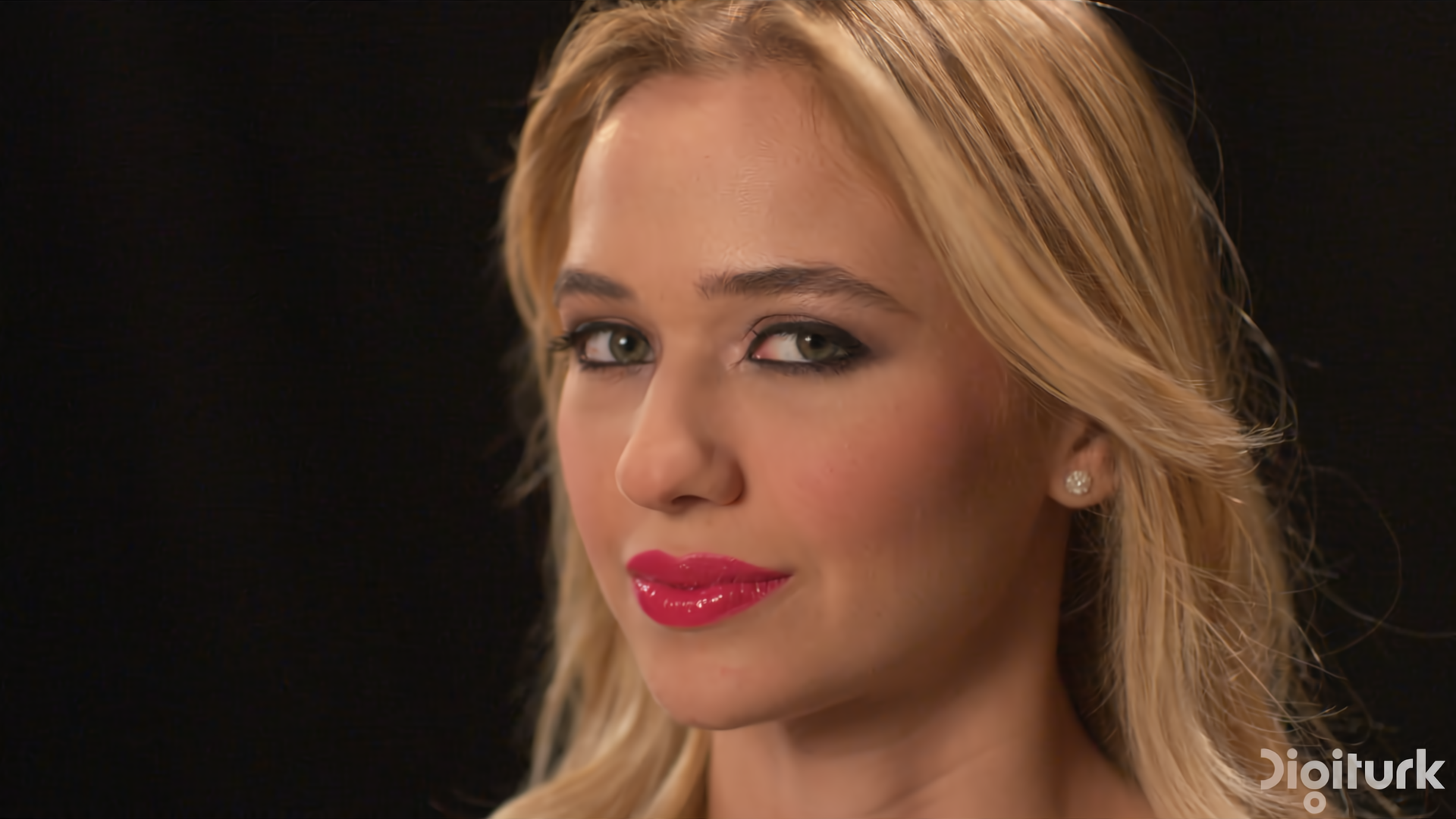}}
 	\subfigure[Proposed (MS-SSIM), 0.0665 bpp, 33.6266 dB, 0.9147]{\includegraphics[scale=0.10]{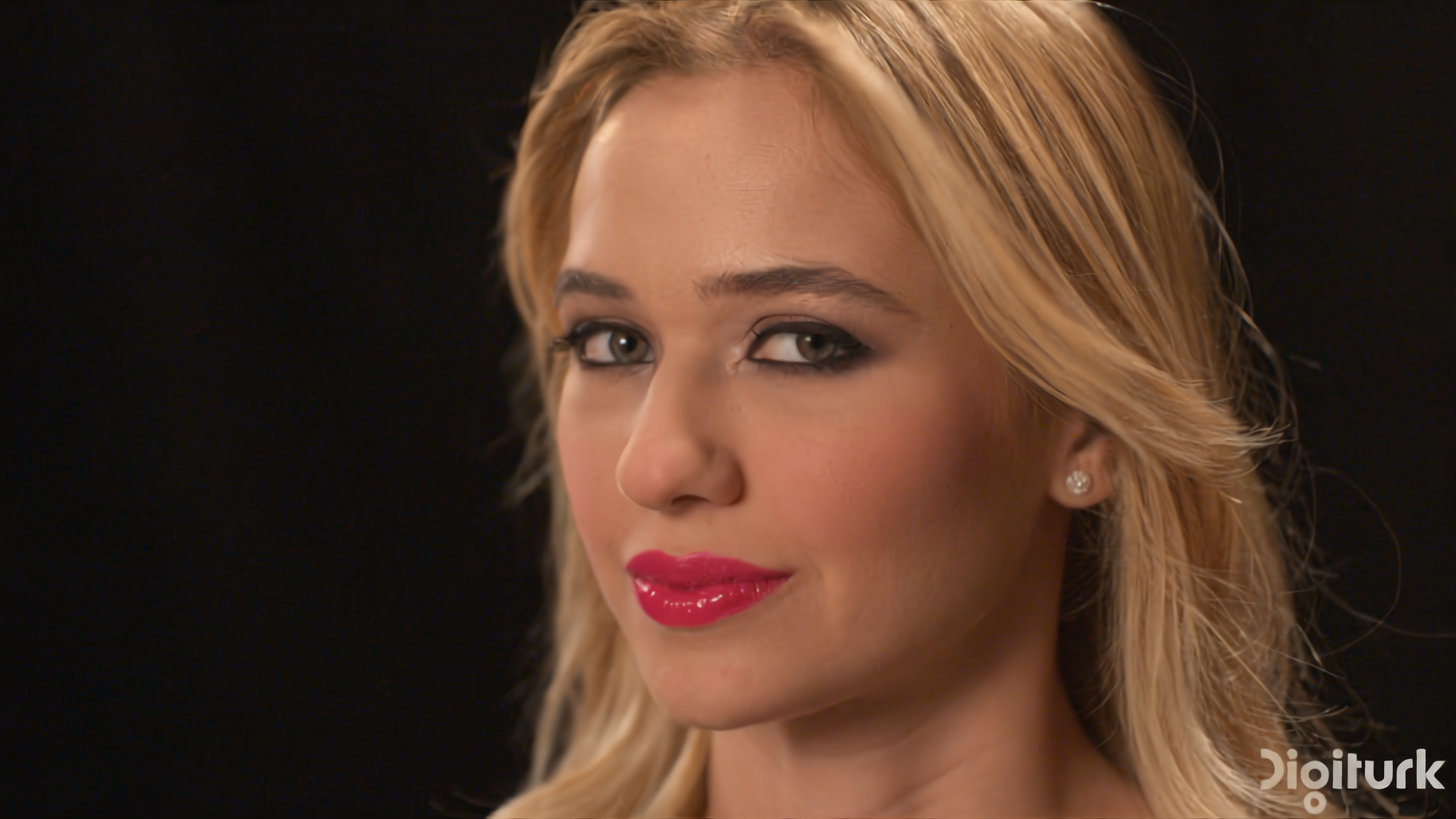}}
 	\caption{Comparison between the proposed method, H.264 and H.265 on the \emph{Beauty} sequences in UVG dataset. Our proposed methods save more bits at the same reconstruction quality level.}
 	\label{fig:BTComp}
 \end{figure*}

 \begin{figure*}[h]
 	\centering
 	\subfigure[H.264, 0.1006 bpp, 34.98 dB, 0.9717]{\includegraphics[scale=0.10]{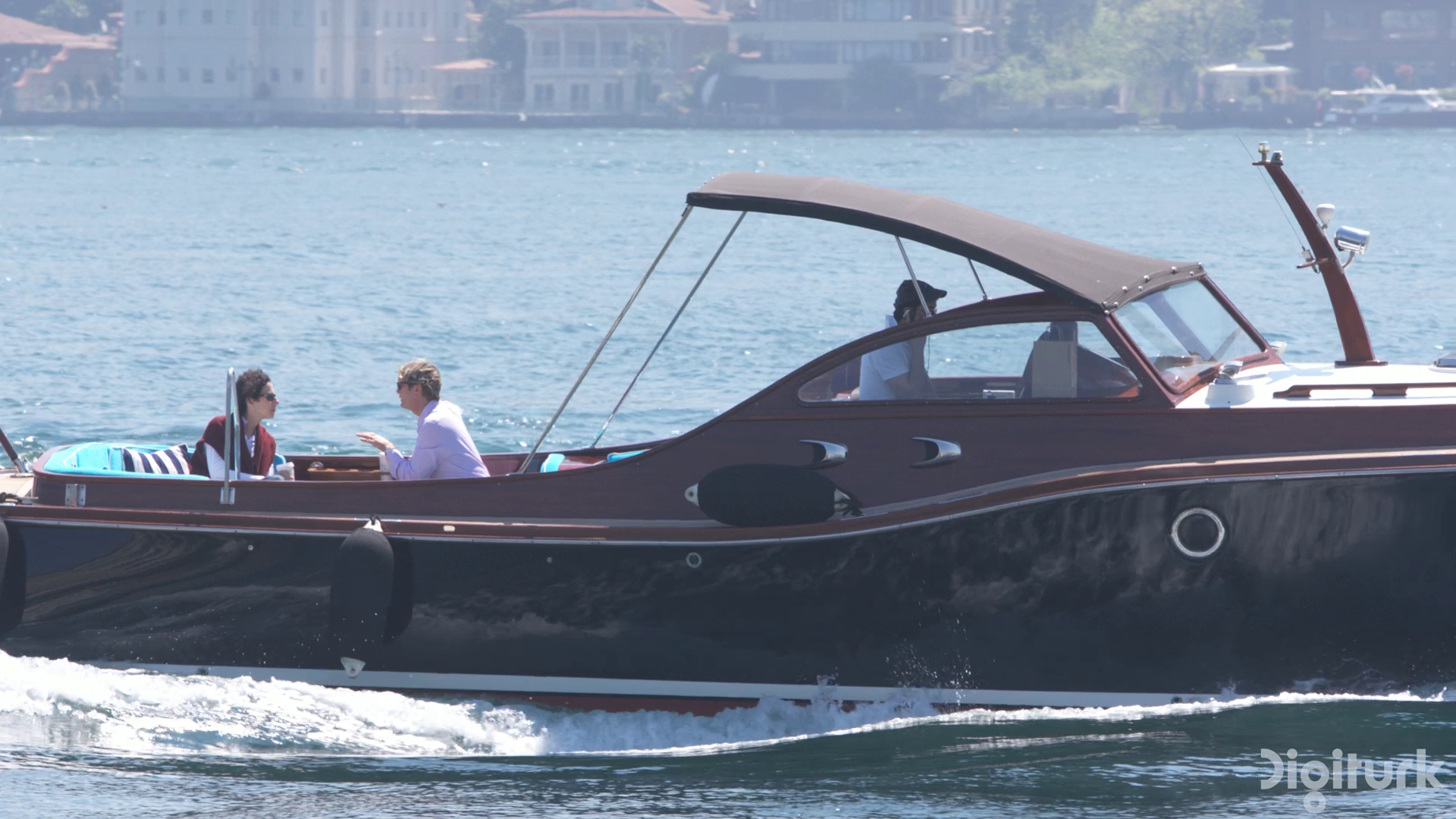}}
 	\subfigure[H.265, 0.1208 bpp, 36.60 dB, 0.9766]{\includegraphics[scale=0.10]{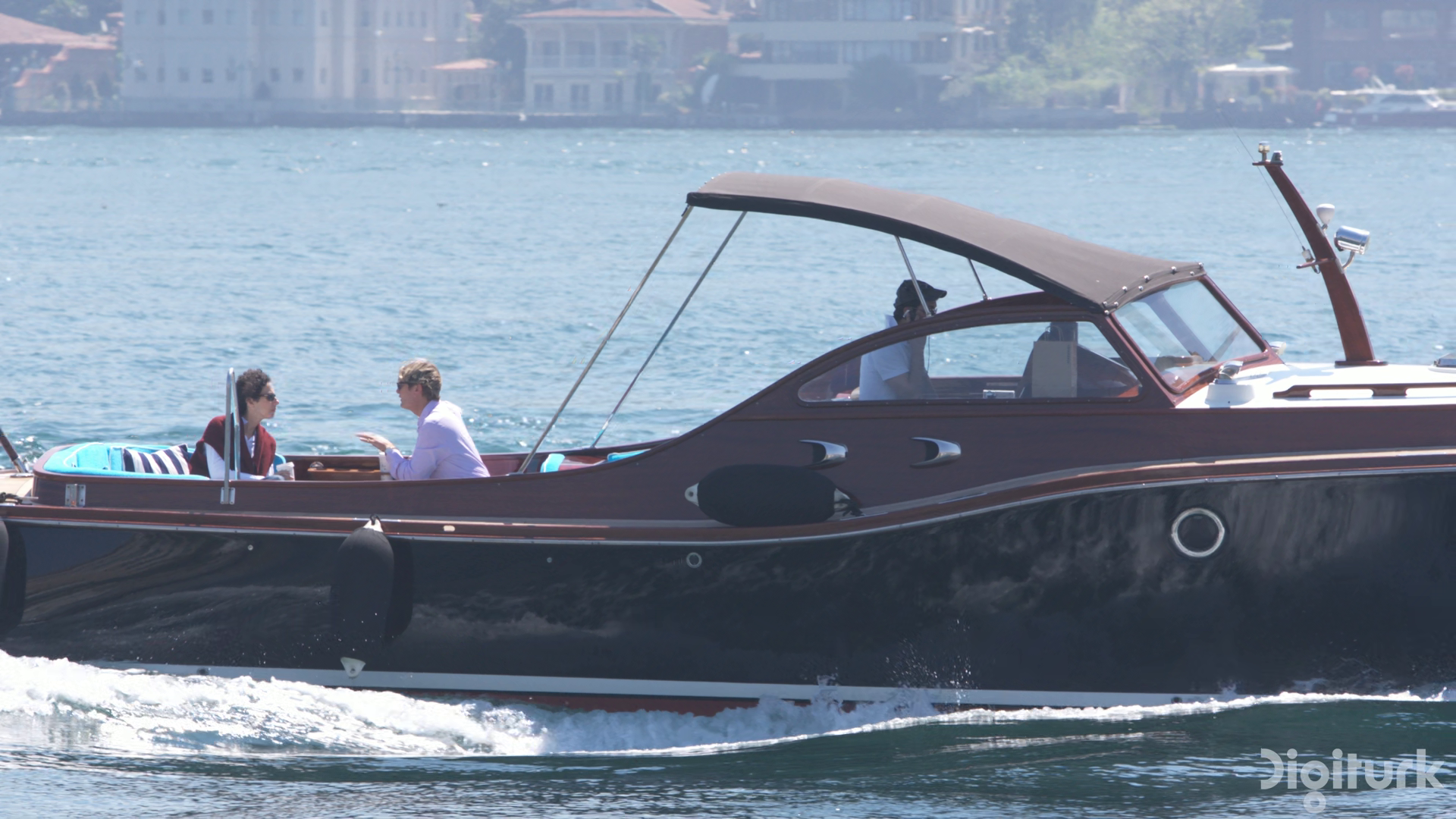}}
 	\subfigure[Proposed (MSE), 0.1082 bpp, 36.77 dB, 0.9742]{\includegraphics[scale=0.10]{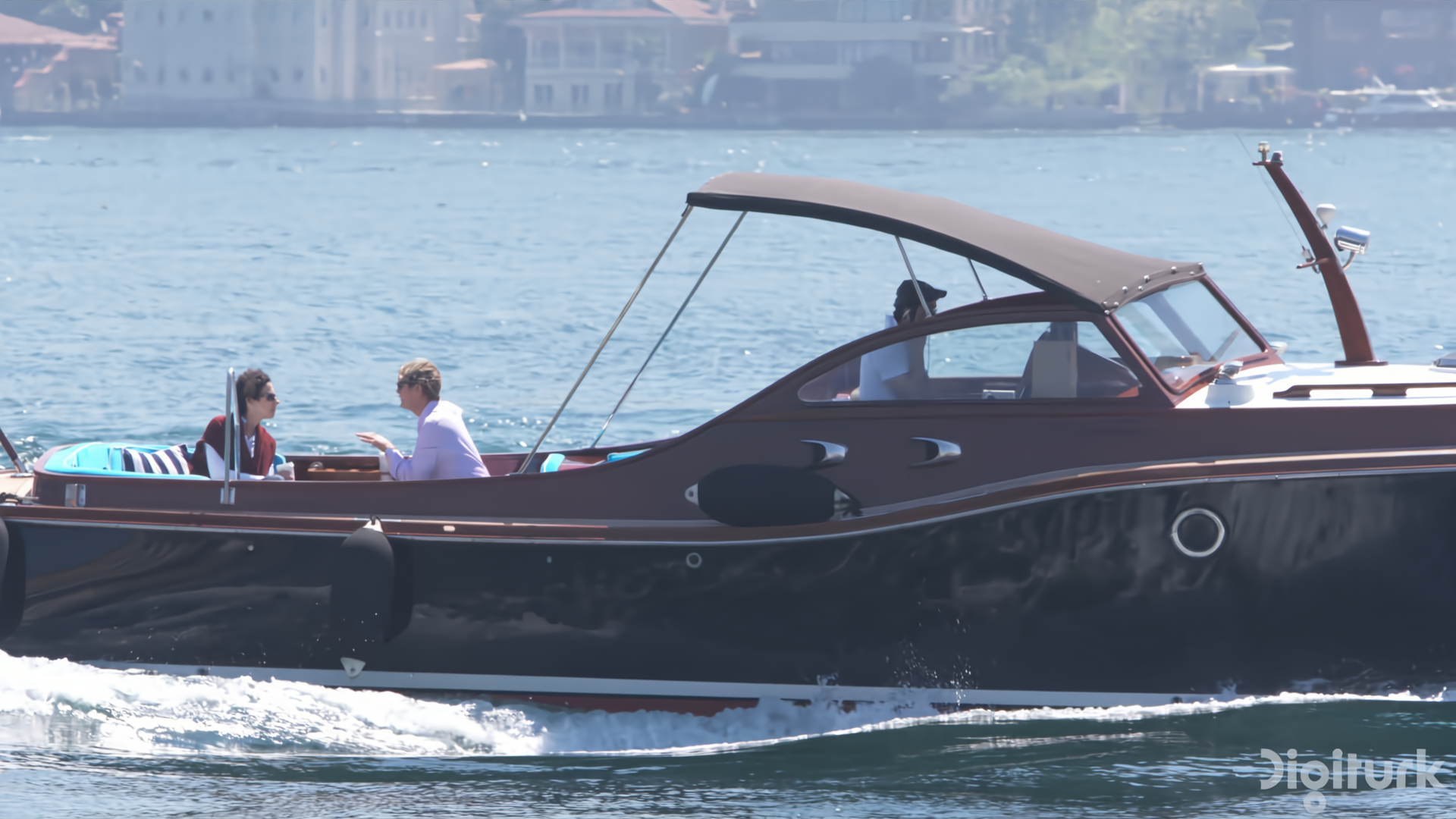}}
 	\subfigure[Proposed (MS-SSIM), 0.0720 bpp, 32.75 dB, 0.9746]{\includegraphics[scale=0.10]{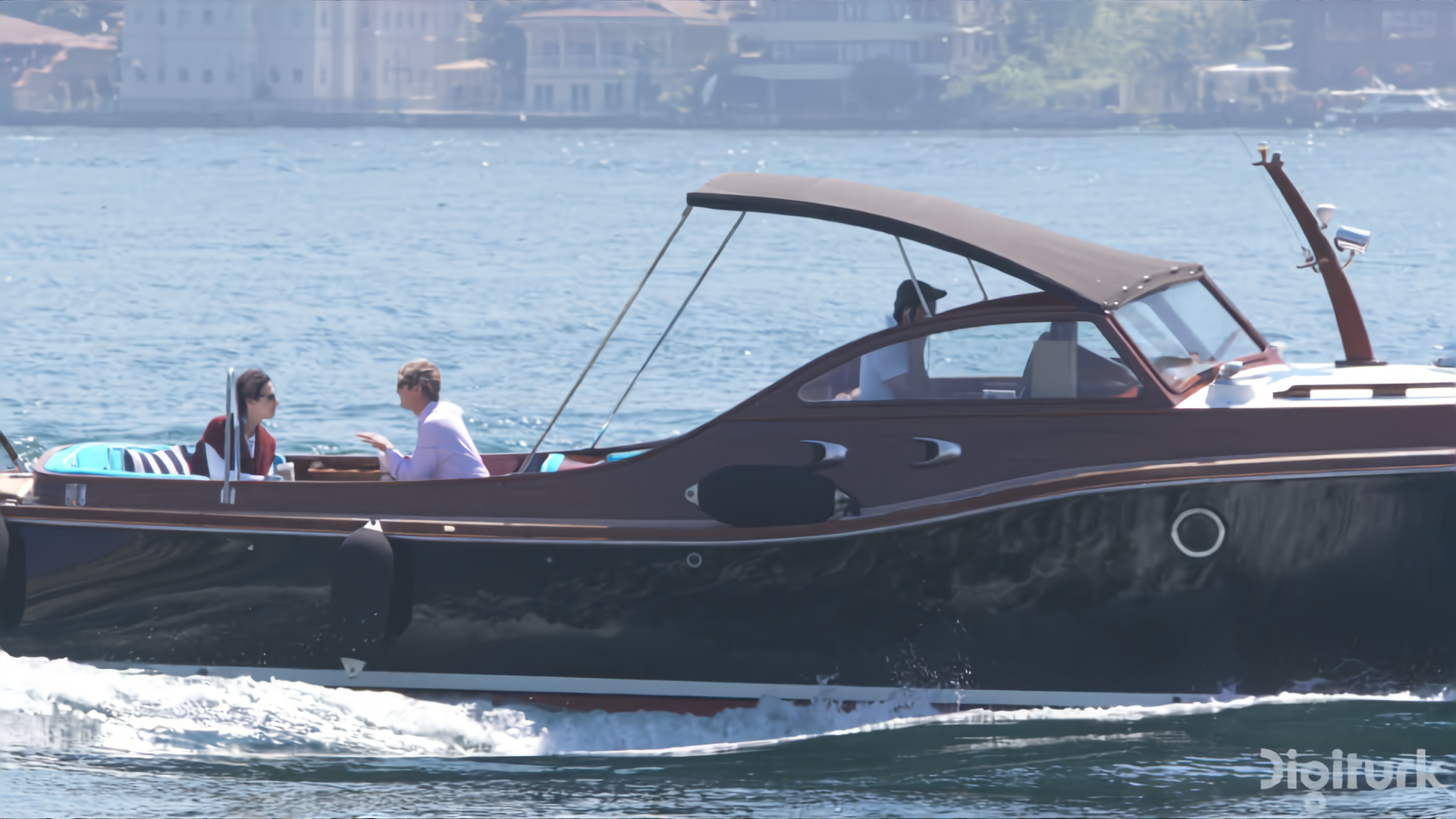}}
 	\caption{Comparison between the proposed method, H.264 and H.265 on the \emph{YachtRide} sequences in UVG dataset. Our proposed methods save more bits at the same reconstruction quality level.}
 	\label{fig:YRComp}
 \end{figure*}

\end{document}